\documentclass[aps,prb,twocolumn,longbibliography%
reprint,
superscriptaddress,
bibnotes,
amsmath,amssymb,
aps,
prb,
]{revtex4-2}

\usepackage{graphicx}
\usepackage{dcolumn}
\usepackage{bm}
\usepackage{hyperref}
\usepackage[normalem]{ulem}

\usepackage{multirow}

\usepackage{physics}

\newcommand{\vbq}{{\vb{q}}}
\newcommand{\vbk}{{\vb{k}}}
\newcommand{\vbr}{{\vb{r}}}

\newcommand{\vbG}{{\vb{G}}}
\newcommand{\vba}{{\vb{a}}}
\newcommand{\vbT}{{\vb{T}}}
\newcommand{\mrm}[1]{\mathrm{#1}}
\newcommand{\kgrid}[1]{$#1$$\times$$#1$$\times$$#1$}

\usepackage{xcolor}

\definecolor{sapgreen}{rgb}{0.31, 0.49, 0.16}
\newcommand{\cfout}[1]{{\color{sapgreen}\sout{#1}}}

\begin{document}
	
	\preprint{APS/123-QED}
	
	\title{Efficient all-electron Bethe-Salpeter implementation using crystal symmetries}
	
	\author{Jörn Stöhler}
	\affiliation{
		Institute for Theoretical Physics, RWTH Aachen University, 52056 Aachen, Germany
	}%
	\affiliation{%
		Peter Grünberg Institut,
		Forschungszentrum Jülich and JARA, 
		52425 Jülich, 
		Germany
	}%

\author{Stefan Blügel}
    \affiliation{
		Institute for Theoretical Physics, RWTH Aachen University, 52056 Aachen, Germany
	}%
	\affiliation{%
		Peter Grünberg Institut,
		Forschungszentrum Jülich and JARA, 
		52425 Jülich, 
		Germany
	}%
 
	\author{Christoph Friedrich}%
 \email{
		c.friedrich@fz-juelich.de
	}
	\affiliation{%
		Peter Grünberg Institut,
		Forschungszentrum Jülich and JARA, 
		52425 Jülich, 
		Germany
	}%

	
	
	
	\date{\today}
	
	\begin{abstract}
		
We describe an all-electron implementation of the Bethe-Salpeter equation (BSE) for the calculation of optical absorption spectra in the full-potential linearized augmented-plane-wave (FLAPW) method.
So far, FLAPW implementations have resorted to a simple plane-wave basis for the bare and screened Coulomb potentials, thereby forgoing the all-electron description to some extent. In contrast, we expand the interaction potentials in the
all-electron mixed basis. The Coulomb divergence is treated analytically, allowing for anisotropic screening.
As in most implementations, the BSE is solved by the diagonalization of a two-particle Hamiltonian matrix, whose dimension is proportional to the number of $\mathbf{k}$ points.
Because of the large number of $\mathbf{k}$ points required to converge the BSE, the resulting matrix becomes large even for small unit cells.
We describe a method that exploits the crystal symmetries to accelerate the construction and diagonalization of the two-particle Hamiltonian.
In particular, we employ group theoretical tools to bring the Hamiltonian into block-diagonal form. Furthermore, it is shown that often only one of the blocks needs to be taken into account for the optical absorption spectrum, leading to a considerable speedup of the diagonalization step.
The code allows for the inclusion of spin-orbit coupling and is parallelized with the possibility of storing the Hamiltonian in distributed memory over many nodes, keeping the memory demand on a single node low.
To validate our implementation, we show optical absorption spectra and report exciton binding energies for bulk Si, LiF, and MoS$_2$. By exploiting the crystal symmetries, we can reduce the dimension of the Hamiltonian matrix of Si by a factor of five, resulting in a theoretical 125-fold speedup for the leading cubic scaling term of the diagonalization. The calculated spectra agree with available theoretical and experimental spectra from the literature. The calculated exciton binding energies of 22~meV and 76~meV for Si and MoS$_2$ are closer to experimental values than in previous BSE studies.
\end{abstract}
	
\maketitle	

\section{Introduction}\label{sec:introduction}
	
The Bethe-Salpeter equation (BSE)
\cite{HedinEquations,Onida1995,Rohlfing1998,Onida2002} enables the accurate computation of optical absorption spectra, electron energy-loss spectra, and exciton binding energies from first principles. It provides a unified description of the response of the interacting many-electron system to the electric field of photons or a beam of electrons, treating plasmonic and excitonic excitations on the same footing. Diagrammatically, the former are described by a summation of Green-function bubbles and the latter by a summation of ladder diagrams to all orders. The BSE also incorporates all possible mixtures of the diagrams, which leads to a mutual renormalization of excitons and plasmons.
	
To date, most implementations of the BSE for periodic systems rely on a plane-wave \cite{Enkovaara_2010,marini_yambo_2009,Giantomassi2011} (or Gaussian \cite{rohlfing_electron-hole_2000}) basis  with a pseudopotential approximation for the Kohn-Sham effective potential. In such an implementation, only the valence electrons are treated explicitly, whereas the contribution of the core electrons to the effective potential is approximated and fixed by the chosen pseudopotential. Furthermore, because of the smoothed form of the potential, the valence wave functions are pseudized close to the atomic nuclei, where they miss the rapid variations of the true energy eigenstates. Likewise, high-lying empty states, relevant for the construction of the screened interaction, differ substantially from true energy eigenstates because the pseudopotential approximation is made to work well only for the valence states, while it becomes less accurate for high-lying states \cite{Friedrich2006}.
	
Full all-electron implementations of the BSE for periodic systems are scarce. 
Implementations based on the LAPW basis set \cite{Puschnigg2002,Wien2k-BSE,Vorwerk_2019} and on the linearized muffin-tin orbitals (LMTO) basis set \cite{Pashov20} have been reported. More recently, an implementation of the BSE based on numerical atomic orbitals has also been introduced \cite{Ruiyi25}. 
Both implementations based on the LAPW basis deviate from a full all-electron description, however, in that they project the (bare and screened) interaction potentials onto an auxiliary plane-wave basis, which cannot resolve the rapid variations close to the atomic nuclei. In our implementation, we do not resort to a plane-wave basis but expand all quantities in the LAPW basis and (in the case of the interaction potentials) its related mixed basis \cite{Kotani2002,Friedrich2010,Betzinger2011}.
	
In a common approach, one reformulates the BSE as an eigenvalue problem with an effective electron-hole Hamiltonian. This Hamiltonian is represented in a basis of electron-hole wave-function products. 
In periodic systems, typically a large number of $\mathbf{k}$ points is required to converge the results, which makes the construction and diagonalization of the dense Hamiltonian matrix computationally expensive even for small systems.
A possibility to reduce the computational cost without sacrificing accuracy is by exploiting crystal symmetries. For example, it is possible to restrict the $\mathbf{k}$ points to an irreducible wedge of the Brillouin zone, which is routine in density-functional theory (DFT) codes. It is also possible in implementations of hybrid functionals \cite{Betzinger2010} and the $GW$ approximation \cite{Friedrich2010}, where an additional extended irreducible wedge has to be introduced because of nested $\mathbf{k}$ summations.
	
Here, we describe our use of crystal and time-reversal symmetries to reduce the cost of BSE calculations. 
First, we speed up the construction of the two-particle Hamiltonian. We reduce the number of entries that need to be calculated explicity, and calculate the rest by applying suitable symmetry transformations. 
Second, we bring the large and dense Hamiltonian into a block-diagonal form by a transformation to a symmetry-adapted basis. Often, only one of the blocks contributes to the spectrum. This significantly accelerates the diagonalization of the Hamiltonian, which is the most time-consuming part of the BSE calculation for large systems (or calculations with dense $\mathbf{k}$ grids). 

The symmetry reduction of the Hamiltonian can be combined with essentially any BSE solution technique. Alternative approaches such as the Lanczos-Haydock technique \cite{Haydock80,Gruening2011} or the time-propagation method \cite{schmidt2003,hahn_quasiparticle_2005} also benefit from the block-diagonal form of the Hamiltonian. However, these techniques only give access to response functions, including the optical absorption spectrum. Individual eigensolutions cannot be obtained without additional algorithms or postprocessing tools. Dark excitons are not directly accessible. Iterative eigensolvers such as the Jacobi-Davidson method \cite{Davidson75,Sleijpen96}, on the other hand, efficiently provide eigensolutions but are not efficient for computing the whole eigenvalue spectrum (and absorption spectrum). Since neither of these alternatives provides access to all eigenstates and the complete absorption spectrum simultaneously, we resort to the standard dense diagonalization in this work.

We would like to emphasize that the symmetry-accelerated construction of the Hamiltonian and its transformation to the symmetry-adapted basis are exact. The computational speedup does not come at the expense of accuracy. (The smaller matrix size may even help reduce numerical rounding errors.)
This stands in contrast to other techniques such as double-grid methods \cite{Rohlfing1998,Alliati21}, which necessarily introduce approximations. The approaches are, of course, compatible and can be combined.
	
In Sec.~\ref{sec:methology}, we  briefly revisit the BSE in its
formulation as an eigenvalue problem. We discuss our choice of basis sets, the LAPW and mixed bases in Sec.\ \ref{sec:implementation}. Then, we describe our implementation in the all-electron code SPEX \cite{Friedrich2010}, in particular the use of crystal symmetries in constructing and diagonalizing the electron-hole Hamiltonian. SPEX is part of the FLEUR family of codes \cite{NIC_FLAPW}. In Sec.~\ref{sec:results}, we present illustrative results 
for Si, LiF, and bulk MoS$_2$ and compare them with the literature. The use of crystal symmetries significantly speeds up the computations, as illustrated by a BSE solution of Si on a very dense 60$\times$60$\times$60 $\vbk$-point grid. Section \ref{sec:conclusions} concludes the paper with a summary of the main results.
The BSE implementation based on group theory has previously been described in Ref.~\onlinecite{stoehler21}, the master’s thesis of the first author.

\section{Theory}\label{sec:methology}
	
\newcommand{\eff}{{\text{eff}}}

Theoretical optical absorption spectra of solids are given by the imaginary part of the macroscopic electronic dielectric function $\varepsilon_\mathrm{M}(\omega)$, which is related to the inverse microscopic electronic dielectric function by \cite{Adler62,Wiser63}
\begin{equation}
	\varepsilon_\mathrm{M}(\omega) = \lim_{\vbq\rightarrow\mathbf{0}} \frac{1}{\varepsilon^{-1}_\mathbf{00}(\mathbf{q},\omega)} \label{eq:def_macro_diel}
\end{equation}
with the plane-wave representation
\begin{equation}
    \varepsilon_\mathbf{GG'}(\mathbf{q},\omega)
	    =\frac{1}{V}\iint e^{i[(\vbq+\vbG')\vbr'
	    -(\vbq+\vbG)\vbr]}
	\varepsilon(\vbr,\vbr';\omega) \,d^3r\,d^3r'
	    \label{eq:micro_diel_expand}
\end{equation}
and the crystal volume $V$.
The photon momenta are small compared to typical electron momenta; hence, the limit $\vbq \to \vb{0}$.
According to Eq.~(\ref{eq:def_macro_diel}), we are not interested in the full inverse matrix but only in the long-wavelength limit of its reciprocal head element (i.e., the $\vbG=\vbG'=\vb{0}$ component), which can be conveniently calculated by
\begin{equation}
	\varepsilon_\mathrm{M}(\omega)=1-\lim_{\vbq\rightarrow \vb{0}}\frac{4\pi}{q^2}\overline{L}_{\vb{0}\vb{0}}(\vbq,\omega).
	\label{eq:macro_diel}
\end{equation}

We note that the right-hand side is a tensor if the material is
anisotropic (bulk MoS$_2$ is an example); i.e., 
the macroscopic dielectric function depends on the direction $\vu{q} = \vb{q} / q$ along which the limit $\vbq\rightarrow\vb{0}$ is taken. This direction corresponds to the light polarization vector. $\varepsilon_\mrm{M}(\omega)$ is then the 3$\times$3 dielectric tensor (and the "1" on the right-hand side the 3$\times$3 identity matrix). If the polarization vector is defined, we know from which direction the limit $\vbq\rightarrow\vb{0}$ is to be taken, and we can interpret Eq.~(\ref{eq:macro_diel}) as a scalar equation.

The respective element of the modified reducible polarizability is given by
\begin{align}
    \overline{L}_{\vb{0}\vb{0}}(\vbq,\omega)
    &\stackrel{\vbq\rightarrow\vb{0}}{\sim}
    \frac{q^2}{N_\vbk}
    \sum_\lambda
    \left|
    B_\lambda(\hat{\vbq})
    \right|^2\nonumber\\
    &\times\left(
    \frac{1}{\omega-\Omega_\lambda+i\eta}-
    \frac{1}{\omega+\Omega_\lambda-i\eta}
    \right)\label{eq:reducible_polar}
\end{align}
with
$N_\vbk$, the number of $\vbk$ points, and the oscillator strength
\begin{equation}
    B_\lambda(\hat{\vbq})=\sum_{\vbk}\sum_{uo}
    A^\lambda_{\vbk uo}B_{\vbk uo}(\hat{\vbq})
    \label{eq:oscillatorA}
    \end{equation}
    and
    \begin{align}
    B_{\vbk uo}(\hat{\vbq})&=\lim_{q\rightarrow 0}\frac{1}{q\sqrt{V_\mathrm{uc}}}
    \langle e^{i\vbq\vbr}\varphi_{\vbk o}|\varphi_{\vbk+\vbq u}\rangle\nonumber\\ 
    &=\frac{1}{\sqrt{V_\mathrm{uc}}}
    \frac{\langle \varphi_{\vbk o} | -i(\hat{\vbq}\nabla) | \varphi_{\vbk u} \rangle}{\epsilon^0_{\vbk u}-\epsilon^0_{\vbk o}}
    \label{eq:oscillator}
\end{align}
where $V_\mathrm{uc}$ is the unit-cell volume.
Note that the factor $q^2$ in Eq.~(\ref{eq:reducible_polar}) cancels out with the factor $1/q^2$ of Eq.~(\ref{eq:macro_diel}). The sum over $o$ and $u$ runs
 over occupied and unoccupied single-particle states, $\varphi_{\vbk o}(\vbr\sigma)$ and $\varphi_{\vbk u}(\vbr\sigma)$, respectively, and $\epsilon^0_{\vbk o}$ and $\epsilon^0_{\vbk u}$ are the corresponding energy eigenvalues of the mean-field system, usually the Kohn-Sham system of DFT.

The spin index $\sigma$ is not treated as a quantum number but is included as an argument to incorporate spin-orbit coupling in the notation. The two components $\{\varphi_{\vbq n}(\vbr\uparrow),\varphi_{\vbq n}(\vbr\downarrow)\}$ thus form spinor wave functions and their inner product is defined by $\langle...\rangle=\sum_\sigma\int_V ... d^3r$. The wave functions are normalized with respect to the whole (infinite) crystal volume $\langle
\varphi_{\vbk n}|
\varphi_{\vbk' n'}\rangle=\delta_{\vbk\vbk'}\delta_{nn'}$.
    
The frequencies $\Omega_\lambda$ in Eq.~\eqref{eq:reducible_polar} and the (normalized) vectors $A^\lambda_{\vbk uo}$ in Eq.~\eqref{eq:oscillatorA} are the eigensolutions of the eigenvalue problem
\begin{align}
    \sum_{\vbk'}\sum_{u'o'}&
    [
    (\epsilon_{\vbk u}
    -\epsilon_{\vbk o})
    \delta_{\vbk \vbk'}
    \delta_{uu'}
    \delta_{oo'}
    +\overline{v}_{\vbk u,\vbk' o';\vbk o,\vbk' u'}
    \nonumber\\
    &-W_{\vbk u,\vbk' o';\vbk' u',\vbk o}
    ]
    A^\lambda_{\vbk' u'o'}
    =
    \Omega_\lambda A^\lambda_{\vbk uo}
    \label{eq:BSE-eigval}
\end{align}
with the single-particle energies $\epsilon_{\vbk n}$. Importantly, we
have to make a distinction between the $\epsilon_{\vbk n}$ of Eq.\ ({\ref{eq:BSE-eigval}}) and the mean-field eigenvalues $\epsilon^0_{\vbk n}$. The latter appear in Eq.\ (\ref{eq:oscillator}) because of the application of $\vbk\cdot\mathbf{p}$ perturbation theory in the last step
of the derivation and must be the exact eigenvalues of the mean-field Hamiltonian with the eigenfunctions $\varphi_{\vbk n}(\vbr)$. The $\epsilon_{\vbk n}$ of Eq.\ (\ref{eq:BSE-eigval}), on the other hand, are the poles of the single-particle Green function, for which one often uses a $GW$-renormalized Green function. In this case, the $\epsilon_{\vbk n}$ would correspond to the $GW$ quasiparticle energies. It is also possible to apply a scissor operator instead. The energies $\epsilon_{\vbk n}$ would then be the "scissored" single-particle energies.
    
The matrix on the left-hand side of Eq.~(\ref{eq:BSE-eigval}) can be understood as an electron-hole Hamiltonian with the interaction matrix elements 
\begin{align}
    &w_{\vbk n,\vbk' n';
    \vbk-\vbq m,\vbk'+\vbq m'}
    =\sum_{\sigma\sigma'}\iint
    \varphi^*_{\vbk n}(\vbr \sigma)
    \varphi^*_{\vbk' n'}(\vbr' \sigma')\nonumber\\
    &\quad\quad\times
    w(\vbr,\vbr')
    \varphi_{\vbk-\vbq m}(\vbr \sigma)
    \varphi_{\vbk'+\vbq m'}(\vbr' \sigma')\,
    d^3r\,d^3r',
    \label{eq:w_interaction}
\end{align}
where the generic interaction potential $w(\vbr,\vbr')$ denotes either the modified bare Coulomb interaction $\overline{v}(|\vbr-\vbr'|)$ or the static screened interaction $W(\vbr,\vbr')$. The former is defined via its Fourier transformation $\overline{v}(\vbq+\vbG)=(1-\delta_{\vbq+\vbG,\vb{0}})4\pi/|\vbq+\vbG|^2$
($\vbG$ are reciprocal lattice vectors), which corresponds to the bare Coulomb interaction without its long-range Fourier component. The elimination of the long-range component enables Eq.~(\ref{eq:macro_diel}) \cite{Onida2002} and is also the reason for the notation $\overline{L}$, which a reminder of this modification. Secondly, the static screened interaction $W(\vbr,\vbr')\equiv W(\vbr,\vbr';\omega=0)$ is calculated within the random-phase approximation. Its dynamic generalization $W(\vbr,\vbr';\omega)$ is routinely used in the $GW$ method \cite{Friedrich2010}.
	
The BSE, written in the form of Eq.~(\ref{eq:BSE-eigval}), can be understood as a stationary Schrödinger equation for electron-hole eigenstates. 
The electron-hole eigensolutions are given by the excitation energies $\Omega_\lambda$ and their eigenvectors  $A^\lambda_{\vbk ou}$.
The effective electron-hole Hamiltonian $H$
acts on the space of two-particle wave functions $\psi(\vbr \sigma, \vbr' \sigma')$, which track the electron and hole positions.
Since the set of single-particle eigenstates $\{\varphi_{\vbk n}(\vbr\sigma)\}$ is complete, it is possible to expand $\psi(\vbr\sigma,\vbr'\sigma')$ in terms of the products $\{\varphi_{\vbk u}(\vbr\sigma)\varphi^*_{\vbk o}(\vbr'\sigma')\}$. In the present case of charge-neutral excitations with a static screened interaction, the indices $u$ and $o$ refer to electron (or unoccupied) and hole (or occupied) states.
In the case of a nonmagnetic system without spin-orbit coupling, the
spin summation amounts to a spin factor of $2$ on the right-hand side of Eq.~(\ref{eq:reducible_polar}) and in front of $\overline{v}$ in Eq.~(\ref{eq:BSE-eigval}).
We use the Tamm-Dancoff approximation, which restricts the eigenvalue spectrum to the so-called resonant electron-hole excitations $ou\leftrightarrow o'u'$ and neglects the coupling to anti-resonant pairs $ou\leftrightarrow u'o'$.

\section{Implementation}
\label{sec:implementation}
	
\subsection{Basis sets}
\label{sub:Basissets}
	
We employ the LAPW basis to represent the single-particle states  $\varphi_{\mathbf{k}n}(\vbr\sigma)$. The FLAPW method\cite{Andersen75,Koelling75,Wimmer81} divides space into two regions, the nonoverlapping atom-centered muffin-tin (MT) spheres and the remaining interstitial region. In the latter, the basis functions are simple plane waves $e^{i(\vbk+\vbG)\vbr}/\sqrt{V}$ with a reciprocal cutoff radius $|\vbk+\vbG|\le G_\mathrm{max}$. These plane waves are matched in value and gradient to linear combinations of numerical functions $u_{lp}(r)Y_{lm}(\hat{\vbr})$ in the MT spheres ($\vbr$ measured from the MT center). Here, $Y_{lm}(\hat{\vbr})$ are spherical harmonics, and $u^\sigma_{lp}(r)$ are numerical functions defined on a radial grid. The $l$ quantum numbers are bounded from above $0\le l\le l_\mathrm{max}$, where $l_{\mathrm{max}}$ is a convergence parameter. The index $p$ enumerates different radial functions. The standard LAPW basis employs $u^\sigma_{l0}(r)$, which is the solution of the radial scalar-relativistic Dirac equation with the spherically averaged spin-$\sigma$ effective Kohn-Sham potential inside the MT sphere, and its energy derivative $u^\sigma_{l1}(r)$. So-called local orbitals \cite{Singh91} $u^\sigma_{lp}(r)$ with $p\ge 2$ can be included to augment the basis in the MT spheres \cite{Singh91}. No approximations are made to the shape of the effective potential \cite{Wimmer81}.
	
As already discussed in the previous section, we formulate the BSE as an eigenvalue problem [Eq.~(\ref{eq:BSE-eigval})] in a two-particle product basis, defined as pairs of occupied and unoccupied states 
\begin{equation}
	    \psi_{\vbk uo}(\vbr\sigma,\vbr'\sigma')=
	    \varphi_{\vbk u}(\vbr\sigma)
	    \varphi^*_{\vbk o}(\vbr'\sigma')~.
	    \label{eq:2p-product-basis}
\end{equation}
Formally, the single-particle wave functions are orthonormal and complete in the electron Hilbert space. This implies that the two-particle product basis is orthonormal and complete in the electron-hole Hilbert space with the orthonormality condition
	\begin{equation}
	    \langle
	    \psi_{\vbk uo}|\psi_{\vbk' u'o'}
	    \rangle=
	    \langle
	    \varphi_{\vbk u}|\varphi_{\vbk' u'}
	    \rangle
	    \langle
	    \varphi_{\vbk o}|\varphi_{\vbk' o'}
	    \rangle=
	    \delta_{\vbk\vbk'}\delta_{uu'}\delta_{oo'}\,.
	\end{equation}
	
As a third basis set, we introduce the mixed basis \cite{Kotani2002,Friedrich2010}, which is designed to represent wave-function products of the form 
$\varphi_{\vbk u}(\vbr\sigma)
\varphi^*_{\vbk' o}(\vbr\sigma)$. In contrast to the electron-hole
basis function of Eq.~(\ref{eq:2p-product-basis}), the two
single-particle wave functions are evaluated at the same point in space
$\vbr$ and with the same spin $\sigma$ (but not necessarily at the
same $\vbk$). The mixed basis is helpful for the evaluation of the
interaction matrix elements of Eq.~(\ref{eq:w_interaction}), in which there are two pairs of wave functions that have the same space and spin arguments. While  the  two-particle  product basis of Eq.~(\ref{eq:2p-product-basis}) is orthonormal, the same-$\vbr$--same-$\sigma$ products are highly linearly dependent, which is why they cannot directly be used as a basis set. Instead, the mixed basis is constructed directly from products of  LAPW basis functions: plane waves $e^{i(\vbk+\vbG)\vbr}/\sqrt{V}$ in the interstitial (with a cutoff $|\vbk+\vbG|\le G'_\mathrm{max}$) matched to linear combinations of $M_{LP}(r)Y_{LM}(\hat{r})$ with $0\le L\le L_\mathrm{max}$. The radial functions $M_{LP}(r)$ are constructed from products $u^\sigma_{lp}(r)u^\sigma_{l'p'}(r)$ with $|l-l'|\le L\le l+l'$. Linear dependencies are eliminated in each $L$ channel to optimize the MT basis. (We note that the mixed-basis functions are defined in a spin-independent manner. In case of spin polarization, both products $u^\uparrow u^\uparrow$ and $u^\downarrow u^\downarrow$ are included in the construction so that not only the products themselves but also differences such as $u^\uparrow u^\uparrow-u^\downarrow u^\downarrow$ or mixed products $u^\uparrow u^\downarrow$ can be well represented in the basis.) 
The mixed-basis functions, denoted by $M^{\vb{q}}_I(\vb{r})$ with an index $I$, are not orthonormal in the interstitial region so that a second dual basis set $\{\tilde{M}^{\vb{q}}_I(\vbr)\}$ needs to be considered. Together they fulfill the completeness relations
	\begin{eqnarray}
		\sum_{\vb{q}} \sum_{I} |M^{\vb{q}}_I\rangle
		\langle\tilde{M}^{\vb{q}}_I| = 1
		\,,\quad
		\langle M^{\vb{q}}_I|\tilde{M}^{\vb{q}^\prime}_J\rangle = \delta_{\vb{q} \vb{q}^\prime} \delta_{I J}\,
		\label{eq:MPB-complete}
	\end{eqnarray}
For more details about the mixed basis, the reader is referred to Refs.~\onlinecite{Kotani2002,Friedrich2010,Betzinger2011}.

		
With the completeness relation Eq.~(\ref{eq:MPB-complete}), the interaction matrix elements Eq.~(\ref{eq:w_interaction}) can be evaluated as vector-matrix-vector products
\begin{align}
    &w_{\vbk n,\vbk' n';
    \vbk-\vbq m,\vbk'+\vbq m'}
    =\frac{1}{N_\vbk}\sum_{IJ}
    \langle \varphi_{\vbk n} | \varphi_{\vbk-\vbq m} M_{\vbq I}\rangle\nonumber\\
    &\quad\quad\quad\times\langle \tilde{M}_{\vbq I}|w|\tilde{M}_{\vbq J}\rangle
    \langle M_{\vbq J} \varphi_{\vbk' n'} | \varphi_{\vbk'+\vbq m'} \rangle\,.
    \label{eq:w_interaction2}
\end{align}
The matrix $\langle \tilde{M}_{\vbq I}|w|\tilde{M}_{\vbq J}\rangle$ is precalculated. 
The prefactor $N_\vbk^{-1}$ originates from the mixed-basis representation $w(\vbr,\vbr')=(N_\vbk)^{-1}\sum_{\vbk,IJ}\langle \tilde{M}_{\vbk I}|w|\tilde{M}_{\vbk J}\rangle M_{\vbk I}(\vbr) M^*_{\vbk J}(\vbr')$.
The computation of the vectors $\langle M_{\vbq I} \varphi_{\vbk n} | \varphi_{\vbk+\vbq m} \rangle$ and the vector-matrix-vector products in Eq.~(\ref{eq:w_interaction2}) are among the most expensive steps of the whole calculation. In Sec.~\ref{subsec:sym_screen}, we will use symmetry operations to accelerate these computations.
    
By calculating the interaction matrix elements using the mixed basis, SPEX maintains the all-electron description and avoids a mapping to a pure plane-wave basis. 
One may wonder why previous FLAPW implementations chose a plane-wave
basis. One reason for this choice might have been the fact that the Coulomb interaction is long-range, which gives rise to singularities in the Coulomb matrix.
The plane-wave representation has the advantage that there is only one divergent element, namely the head element ($\vbG=\vbG'=\vb{0}$), in the bare Coulomb matrix, and the screened interaction additionally has divergent wing elements ($\vbG=\vb{0}$ or $\vbG'=\vb{0}$) of lower order. This simple structure is lost in the mixed-basis representation. 
However, it can be recovered if a unitary transformation from the mixed basis to the Coulomb eigenbasis is performed \cite{friedrich_efficient_2009}. The mixed basis is defined in such a way that this transformation leads to the same matrix structure as in the pure plane-wave basis with divergent head and wing elements, while all other matrix elements remain finite. We can thus treat the Coulomb singularity as if we had a pure plane-wave basis. Details are discussed in Appendix \ref{sec:appendix}.
 
\subsection{Usage of symmetries}
\label{sec:Symmetries}
	

BSE calculations are computationally demanding. The most expensive steps are (1) the calculation of the screened interaction within the random-phase approximation, (2) the construction of the electron-hole Hamiltonian Eqs.~(\ref{eq:BSE-eigval}) and (\ref{eq:w_interaction}), and (3) the diagonalization of the electron-hole Hamiltonian. All of these steps can be significantly accelerated by exploiting spatial and time-reversal symmetries, which is an elegant way of reducing the computational burden without sacrificing accuracy.


How spatial and time-reversal symmetries can be utilized in the calculation of the screened interaction has already been described in Ref.\ \onlinecite{Friedrich2010}. For the sake of completeness, we note that we make use of the block-diagonal structure of $W$ in the basis of Coulomb eigenfunctions $\{C_{\vbk\mu}\}$: $\langle C_{\vbk\mu}|W|C_{\vbk\nu}\rangle=0$ unless $C_{\vbk\mu}(\vbr)$ and $C_{\vbk\nu}(\vbr)$ fall into the same irreducible representation (irrep) of group theory.  
The Coulomb eigenfunctions are defined as linear combinations of the mixed-basis functions $C_{\vbq\mu}(\vbr)=\sum_I c_{\vbq,\mu I}M_{\vbq I}(\vbr)$ with the matrix $c_{\vbq,\mu I}$ of eigenvectors of $v_{\vbq,I J}=\langle M_{\vbq I}|v|M_{\vbq J}\rangle$. Obviously, $\overline{v}$ becomes diagonal in this basis. 
We will discuss irreps in Sec.\ \ref{sub:Irreps} in more detail. Presently, it suffices to understand the irreps as a classification scheme that allows us to bring the matrix representation of $W$ into block-diagonal form. 
A similar technique can be used for nonlocal operators \cite{Betzinger2010} in general, including the $GW$ self-energy.
The block-diagonal structure of $W$ accelerates the evaluation of matrix-matrix products. However, since this aspect is not  the focus of the present paper, we defer a more detailed discussion to future work.

	
	
The evaluation of the matrix elements of Eq.\ (\ref{eq:w_interaction2}) is the most expensive step in the construction of the electron-hole Hamiltonian of Eq.\ (\ref{eq:BSE-eigval}).
Fortunately, many of the $W$ matrix elements are related by symmetry. Rather than calculating all matrix elements explicitly, we compute only a representative subset and generate the remaining elements by applying symmetry transformations. This subset is evaluated for $\vbk$ and $\vbk'$ belonging to different wedges of the Brillouin zone.

Furthermore, we apply 
group theoretical tools to make the two-particle Hamiltonian
block-diagonal and, in this way, speed up the diagonalization of the
Hamiltonian. It is important to note that, in this case, the
Hamiltonian is a two-particle Hamiltonian. It is represented in a
basis of two-particle functions, each depending on two points in space
(see Sec.\ref{sub:Basissets}). So, the block-diagonality of the
electron-hole Hamiltonian goes beyond what was explained above about
the $W$ matrix, which is represented in mixed-basis functions, i.e.,
regular local functions that depend on a single point in space. To
bring the Hamiltonian into block-diagonal form, we have to perform a
unitary transformation of the product functions. The new functions --- linear combinations of the basis functions in Eq.\ (\ref{eq:2p-product-basis}) --- can then be classified uniquely into irreps.


To the best of our knowledge, existing BSE implementations exploit spatial symmetries by restricting the $\vbq$ vector of the screened Coulomb interaction $W(\vbq)$ to the irreducible wedge of the Brillouin zone (IBZ) and applying symmetry transformations whenever quantities outside the IBZ are needed. However, we are not aware of a published implementation that reduces the set of explicitly evaluated BSE interaction matrix elements to those with $\vbk$ and $\vbk'$ belonging to irreducible wedges of the Brillouin zone and reconstructs the full kernel by applying symmetry operations directly to the matrix elements. 
Tabulated irreducible representations have recently been used to analyze the symmetry properties of excitonic excitations  \cite{Nalabothula2025}, but without applying them to the BSE method itself, for example to block-diagonalize the Hamiltonian as done in the present work.
All symmetry-based computational measures used to accelerate the calculations are independent of the choice of basis set. In particular, they do not depend on the all-electron LAPW basis set used in the present work.

\subsubsection{\label{subsec:sym_mat}Symmetry transformation matrix}

Before explaining the techniques in detail, we introduce the notation. We write a crystal symmetry operation as $s=(A,\vbT+\vba,\alpha)$ with a $3\times 3$ (proper or improper) rotation matrix $A\in \mathrm{O}(3)$, a lattice vector $\vbT$, a translation vector $\vba$ (which is not a lattice vector), and $\alpha$, which is 0 (1) if the symmetry operation excludes (includes) time reversal. The complete set of symmetry operations forms a symmetry group $S$.

The action of a symmetry operation on a real-space and reciprocal-space vector as well as on a single-particle function $\varphi(\vbr\sigma)$ is declared as
\begin{eqnarray}
s\vbr&=&A\vbr+\vbT+\vba \\
s\vbk&=&(-1)^{\alpha}A\vbk+\overline{\vbG} \label{eq:sk}\\
s\varphi(\vbr\sigma)&=&\sum_{\sigma'} B_{\sigma\sigma'}\, c^{\alpha}\,\varphi(s^{-1}\vbr,\sigma')
\label{eq:sphi}
\end{eqnarray}
with $s^{-1}\vbr=A^{-1}(\vbr-\vbT-\vba)$ and the conjugation operator
$c$ [$c^0f(\vbr)=f(\vbr)$ and $c^{1}f(\vbr)=f^*(\vbr)$]. The vector $\overline{\vbG}$ (if not the null vector) folds the $\vbk$ vector back into the first Brillouin zone. Obviously, $\overline{\vbG}$ depends on $s$ and $\vbk$ but has been written without these dependencies to simplify the notation. The matrix $B\in\mathrm{SU}(2)$ is  the $2\times 2$ rotation matrix in spin space, which derives from $A$ and $\alpha$. If spin-orbit coupling is neglected, the $B$ matrix is simply the identity matrix.


When we let a symmetry operation $s$ that leaves the Hamiltonian invariant act on an eigenstate, it maps this eigenstate onto another eigenstate with the same energy eigenvalue.
The symmetry transformation, when represented in the basis of eigenstates, is thus a sparse unitary matrix, which only mixes degenerate states
\begin{eqnarray}
s\,\varphi_{\vbk n}(\vbr\sigma) = \sum_{n^\prime\sim n} \Gamma(s)_{\vbk n}^{\vbk' n'}\,\varphi_{\vbk'n^\prime}(\vbr\sigma),
\label{eq:Ts}
\end{eqnarray}
where $\vbk'=s\vbk$ and the notation $n'\sim n$ means that $n'$ runs over all states at $s\vbk$ [Eq.~(\ref{eq:sk})] that are (energy) degenerate with the $n$th state at $\vbk$.
It suffices to calculate $\Gamma(s)_{\vbk n}^{\vbk' n'}=\langle \varphi_{\vbk' n'}|s\varphi_{\vbk n}\rangle$ for the generators $\{s_\nu\}$ of $S$. The generators are elements of $S$ from which all symmetry operations can be generated: For any $s\in S$, there is a representation $s=s_\nu s_{\nu'} s_{\nu''}\cdots$ (some of the generators may be identical). The transformation $\Gamma(s)$ can then be constructed by multiplication $\Gamma(s)=\Gamma(s_\nu)\Gamma(s_{\nu'})\Gamma(s_{\nu''})\cdots$ (in simplified notation). The number of generators of $S$ is usually much smaller than the total number of operations in $S$ (e.g., three generators for 48 symmetry operations in the case of silicon).

Applying the symmetry transformation to an electron-hole basis function yields
	\begin{eqnarray}
		s\,[\varphi_{\vb{k} u}(\vbr\sigma) \varphi_{\vb{k} o}^*(\vbr'\sigma')] &=& 
		\sum_{u' \sim u} \,\sum_{o' \sim o}\,
		\tilde{\Gamma}(s)_{\vbk uo}^{\vbk'u'o'} \\
		\,&\times&\varphi_{\vb{k}'u'}(\vbr\sigma) \varphi_{\vb{k}'o'}^*(\vbr'\sigma')\nonumber
	\end{eqnarray}
again with a sparse unitary transformation matrix
	\begin{eqnarray}\label{eq:2particle-symrep}
	    \tilde{\Gamma}(s)_{\vbk uo}^{\vbk'u'o'} = 
	    \Gamma(s)_{\vbk u}^{\vbk'u'} 
	    \Gamma^*(s)_{\vbk o}^{\vbk'o'}\,.
        \label{eq:Ttilde}
	\end{eqnarray}

The full symmetry group $S$ has thus representations $\tilde{\Gamma}(s)$ of unitary and anti-unitary operators in the space of electron-hole wave-function products. The translations $s_\vbT:\vbr\mapsto\vbr+\vbT\in S_\vbT$
have a trivial representation
\begin{equation}
\tilde{\Gamma}(s_{\vbT})^{\vbk u'o'}_{\vbk u o} = \delta_{u u'} e^{i \vbT \vb{k}} 
\delta_{o o'} e^{-i \vbT \vb{k}}
= \delta_{u u'}
\delta_{o o'}\,.
\end{equation}
We can therefore switch from $S$ to the group quotient $S/S_\vbT$, which removes the lattice translations. The resulting quotient group is now finite and the representations $\tilde{\Gamma}(sS_\vbT)$ of the coset $sS_\vbT$ are simply $\tilde{\Gamma}(s)$. For simplicity, we will refer to the quotient group as $S$ in the following.

\subsubsection{\label{subsec:sym_screen}Hamiltonian matrix elements}

The effective electron-hole Hamiltonian matrix and its elements must obey the crystal symmetries. 
This fact can be used to accelerate its construction. We first turn to the screened interaction matrix $W$. 
Let us consider a matrix element $W_{\vbk' u,\vbk o';\vbk u',\vbk' o}$
(note the change of notation $\vbk\leftrightarrow\vbk'$ with respect
to Eq.~\ref{eq:BSE-eigval}). Comparison with
Eq.~(\ref{eq:w_interaction}) gives $\vbk'=\vbk+\vbq$ and shows that
the $w$ matrix (here, $w=W$) depends on $\mathbf{q}$. The loop over
$\vbq$ should  be the outer loop. We restrict $\vbq$ to the
irreducible Brillouin zone (IBZ), which is the minimal set of $\vbk$ points from which all other $\vbk$ points can be generated via Eq.~(\ref{eq:sk}). In addition, we restrict $\vbk$ to the \emph{extended irreducible Brillouin zone} [EIBZ($\vbq$)], which is the minimal set of $\vbk$ points from which all other $\vbk$ points can be generated via Eq.~(\ref{eq:sk}) with the restriction that the operations $s$ are elements of the so-called "little group". The little group is a subgroup of $S$ that  contains all symmetry operations $s$ that map $\vbq$ onto itself, $s\vbq=\vbq$. In general, the EIBZ($\vbq$) is larger than the IBZ because the little group is smaller than the full symmetry group. 

We now show that all other matrix elements (with $\vbq\in\mrm{BZ}$ and $\vbk\in\mrm{BZ}$) can be generated from the subset [$\vbq\in\mrm{IBZ}$ and $\vbk\in\mrm{EIBZ(\vbq)}$] by applying suitable symmetry operations. Given general $\vbq$ and $\vbk$, there is a symmetry operation $s'$ that rotates $\vbq$ into the IBZ, $\vbq'=s'\vbq$, and there is another symmetry operation $s''$ of the little group that rotates $s'\vbk$ into the EIBZ($\vbq'$). By definition of the little group, $s''$ leaves $\vbq'$ invariant, and we have $\vbq'=s\vbq\in\mrm{IBZ}$ and $\vbk'=s\vbk\in\mathrm{EIBZ(\vbq')}$ with $s=s''s'$. 

Let us write an arbitrary matrix element as
$
\langle 
\varphi_{\vbk+\vbq u}
\varphi_{\vbk o'}|W|
\varphi_{\vbk u'}
\varphi_{\vbk+\vbq o}
\rangle
$. Substitution of the integration variables $\vbr\rightarrow s^{-1}\vbr$ (likewise for $\vbr'$) does not change the integral. Then, using Eq.~(\ref{eq:sphi}), the symmetry invariance of $W(\vbr,\vbr')$ ($s\,W\,s^\dagger=W$), and the fact that $W(\vbr,\vbr')$ is real, we can write
\begin{eqnarray}
\lefteqn{
\langle
\varphi_{\vbk+\vbq u}
\varphi_{\vbk o'}|W|
\varphi_{\vbk u'}
\varphi_{\vbk+\vbq o}
\rangle}\label{eq:Wtrafo}\\
&=&
c^\alpha
\langle 
s\varphi_{\vbk+\vbq u}\,
s\varphi_{\vbk o'}|W|
s\varphi_{\vbk u'}\,
s\varphi_{\vbk+\vbq o}
\rangle\nonumber\\
&=&
c^\alpha
\sum_{\substack{u''\sim u\\u'''\sim u'}}
\sum_{\substack{o''\sim o\\o'''\sim o'}}
\Gamma^*(s)_{\vbk+\vbq u}^{\vbk'+\vbq' u''}
\Gamma^*(s)_{\vbk o'}^{\vbk' o'''}
\Gamma(s)_{\vbk u'}^{\vbk' u'''}\nonumber\\
&&\Gamma(s)_{\vbk+\vbq o}^{\vbk'+\vbq' o''}
\langle 
\varphi_{\vbk'+\vbq' u''}
\varphi_{\vbk' o'''}|W|
\varphi_{\vbk' u'''}
\varphi_{\vbk'+\vbq' o''}
\rangle\nonumber
\end{eqnarray}
with the irreducible representations of Eq.~(\ref{eq:Ts}). We have taken into account the possibility of $\alpha=1$, i.e., the symmetry operation $s$ involves time reversal. The spin rotation matrix $B_{\sigma\sigma'}$ can be ignored because the integrals of the form $\langle\varphi|\varphi M\rangle$ in Eq.~(\ref{eq:w_interaction}) involve a trace over the spins (in the SOC case), which is invariant with respect to SU(2) spin rotations.
Since the four summations only run over the degenerate subspaces (including a maximum of three states), Eq.~(\ref{eq:Wtrafo}) requires practically no computing time. 

In practice, we loop over $\vbq'\in\mrm{IBZ}$ and $\vbk'\in\mrm{EIBZ(\vbq')}$ and calculate the corresponding matrix elements of the right-hand-side of Eq.~(\ref{eq:BSE-eigval}), which we call the "seed" matrix elements. For each pair $(\vbk',\vbk'+\vbq')$, we determine a minimal set of symmetry operations $\{s_1,s_2,\dots\}$ with which all pairs $(\vbk_1,\vbk_2)$ symmetry equivalent to $(\vbk',\vbk'+\vbq')$ can be generated, e.g., $\vbk_1=s_1 \vbk'$ and $\vbk_2=s_1 (\vbk'+\vbq')$. Here, "minimal" means that if there is another $s$ that would generate the same pair, this $s$ is discarded from the set. The set does \emph{not} form a subgroup, since, for example, the neutral element is not an element of the set. 
We then loop over $s_1$, $s_2$, etc.~and calculate the corresponding "rotated" matrix elements following Eq.~(\ref{eq:Wtrafo}). In this way, all matrix elements of the screened Coulomb interaction are calculated. 

An analogous strategy is applied to the 
matrix elements of the (modified) bare Coulomb interaction
$\overline{v}_{\vbk u,\vbk' o';\vbk o,\vbk' u'}$. Furthermore, it is
possible to accelerate the computation of the seed matrix elements.
The different order of $\vbk$ indices compared to $W_{\vbk' u,\vbk
o';\vbk u',\vbk' o}$ leads us to another strategy for evaluating them, namely one that minimizes the number of integrals $\langle M \varphi|\varphi \rangle$ and matrix-vector products.
A comparison to Eq.~(\ref{eq:w_interaction2}) (now, $w=\overline{v}$) shows that $\vbq=\vb{0}$. 
Only a limited number of seed matrix elements $\overline{v}_{\vbk u,\vbk' o';\vbk o,\vbk' u'}$ is required as in the case of $W_{\vbk' u,\vbk o';\vbk u',\vbk' o}$, now with the restriction $\vbk\in\mrm{IBZ}$ and $\vbk'\in\mrm{EIBZ(\vbk)}$. For the sake of generality \footnote{To validate the symmetry implementation, it may be helpful to compare against results obtained for the general case with $\vbk\in\mrm{BZ}$ and $\vbk'\in\mrm{BZ}$.}, we 
adopt in the following the more general assumption $\vbk\in\mrm{BZ}$ and $\vbk'\in\mrm{BZ}$, which is straightforwardly adapted to the simpler case $\vbk\in\mrm{IBZ}$ and $\vbk'\in\mrm{EIBZ(\vbk)}$. First, we calculate the integrals $\langle M_{\vb{0}I}\varphi_{\vbk n}|\varphi_{\vbk m}\rangle$ for all $\vbk\in\mrm{IBZ}$ and multiply them with the Coulomb matrix
\begin{equation}
[ \tilde{M}_{\vb{0}I} \varphi_{\vbk n}|\varphi_{\vbk m} ]
:=\sum_J\langle \tilde{M}_{\vb{0}I} |\overline{v}|\tilde{M}_{\vb{0}J}\rangle \langle M_{\vb{0}J}\varphi_{\vbk n}|\varphi_{\vbk m}\rangle .
\end{equation}
Here, we can make use of a sparse representation of the $\overline{v}$ matrix \cite{Betzinger2010}. Obviously, the matrix elements for $\vbk\in\mrm{IBZ}$ and $\vbk'\in\mrm{IBZ}$ can be obtained directly by scalar products
\begin{equation}
\overline{v}_{\vbk u,\vbk' o';\vbk o,\vbk' u'}=\sum_I \langle \varphi_{\vbk u}|\varphi_{\vbk o}M_{\vb{0}I}\rangle 
[ \tilde{M}_{\vb{0}I} \varphi_{\vbk' o'}|\varphi_{\vbk' u'} ]
\label{eq:v_scalar}
\end{equation}
For $\vbk$ outside the IBZ, the wave functions $\varphi_{\vbk n}(\vbr\sigma)$ are not stored in computer memory, but they are generated by the operation $\varphi_{\vbk n}(\vbr\sigma)=s_\vbk\varphi_{\vb{p}_\vbk n}(\vbr\sigma)$ according to Eq.~(\ref{eq:sphi}) with a suitable operation $s_\vbk$ and $\vb{p}_\vbk\in\mrm{IBZ}$ ($\vb{p}_\vbk$ is the "parent" of $\vbk$ with $s_\vbk \vb{p}_\vbk=\vbk$). With an analogous notation as the one used above for the $W$
matrix elements, we thus have
\begin{eqnarray}
\lefteqn{\langle 
\varphi_{\vbk u}
\varphi_{\vbk o}|\overline{v}|
\varphi_{\vbk' o'}
\varphi_{\vbk' u'}
\rangle}\\
&=&
\langle 
s_\vbk\varphi_{\vb{p}_\vbk u}
s_\vbk\varphi_{\vb{p}_\vbk o}|\overline{v}|
s_{\vbk'}\varphi_{\vb{p}_{\vbk'} o'}
s_{\vbk'}\varphi_{\vb{p}_{\vbk'} u'}
\rangle\nonumber\\
&=&
c^\alpha
\langle
\varphi_{\vb{p}_\vbk u}
\varphi_{\vb{p}_\vbk o}|\overline{v}|
s\varphi_{\vb{p}_{\vbk'} o'}
s\varphi_{\vb{p}_{\vbk'} u'}
\rangle\,,\nonumber
\end{eqnarray}
where $s=s^{-1}_\vbk s_{\vbk'}$, and $\alpha=1$ takes into account time reversal in the symmetry operation $s_\vbk$, otherwise $\alpha=0$. We have used that $\overline{v}(\vbr,\vbr')$ is real and invariant with respect to all $s$ ($s\overline{v}s^\dagger=\overline{v}$). The symmetry transformation has to be applied only to the wave functions on the right-hand side of the equation. We do this by transforming the known $[ \tilde{M}_{\vb{0}I} \varphi_{\vb{p}_{\vbk'} o'}|\varphi_{\vb{p}_{\vbk'} u'} ]$ to $[ \tilde{M}_{\vb{0}I} s\varphi_{\vb{p}_{\vbk'} o'}|s\varphi_{\vb{p}_{\vbk'} u'} ]$. We let $s^{-1}$ act on all quantities in $[...]$, which corresponds to a change of integration variables. The transformation can then be performed in the space of the mixed-product basis ($I$ index), giving $c^{\alpha'}[(s^{-1} \tilde{M}_{\vb{0}I}) \varphi_{\vb{p}_{\vbk'} o'}|\varphi_{\vb{p}_{\vbk'} u'} ]$ with $\alpha'=1$ if $s$ involves time reversal, otherwise $\alpha'=0$.

It seems now that a very large number of such transformations have to be carried out because of the dependence of $s=s^{-1}_\vbk s_{\vbk'}$ on $\vbk\in\mrm{BZ}$ and $\vbk'\in\mrm{BZ}$. Note that the number of $\vbk$ points can be very large in BSE calculations. Of course, the number of symmetry operations for different combinations of $\vbk$ and $\vbk'$ cannot be larger than the total set of symmetry operations. It is therefore sufficient to perform the transformation for all $s\in S$ 
and store the results. The calculation of the matrix elements of Eq.~(\ref{eq:v_scalar}) then amounts to simple scalar products of vectors stored in memory.
	

\subsubsection{Irreducible representations}
\label{sub:Irreps}

One of the computationally most expensive steps in the BSE calculation
is the diagonalization of the electron-hole Hamiltonian; note that the
dimension of the Hamiltonian matrix is large and it grows with $N_\mathrm{o}N_\mathrm{u}N_\vbk$ where $N_\mathrm{o}$, $N_\mathrm{u}$, and $N_\vbk$ are the numbers of occupied and unoccupied bands as well as the number of $\vbk$ points.
The computational cost thus grows cubically with the number of $\vbk$ points, whereas the construction of the screened interaction $W$ and the electron-hole Hamiltonian $H$ exhibit only a quadratic scaling.
The computation time can be reduced if the electron-hole Hamiltonian is brought into block-diagonal form. To achieve this, one has to construct a symmetry-adapted basis, in which the Hamiltonian acquires the desired form. 
This is indeed possible in the present case of a four-point operator, which might be surprising given the fact that group theory is usually applied in theoretical solid-state theory to mathematically less complex matrices, such as the phonon dynamical matrix or single-particle Hamiltonian.

A symmetry-adapted basis can be generated once we know the irreducible representations (irreps) of the symmetry operations (or their traces). As the name suggests, the irreps are the smallest possible matrix representations of the symmetry operations. Equation (\ref{eq:2particle-symrep}) defines valid representations, which, moreover, are already quite sparse, but they are \emph{not} the smallest possible representations. In other words, the representations are reducible. So, the electron-hole products of Eq.\ (\ref{eq:2p-product-basis}) are not yet the symmetry-adapted basis we are looking for. A simple way to see this is that the $\tilde{\Gamma}$ matrix of Eq.~(\ref{eq:2particle-symrep}) would form a $9\times 9$ matrix in the case of threefold degeneracies in the occupied ($\vbk o$) and unoccupied ($\vbk u$) states, but the maximum irrep dimension in space groups is three. Before we explain how a symmetry-adapted basis can be constructed, we have to introduce some concepts of group theory and start by restating the great orthogonality theorem.

	
	
	

We restrict ourselves to linear symmetry operations (excluding time-reversal symmetries) for simplicity.	
Let $S$ be a finite symmetry group, let $\mathcal{H}$ be a finite dimensional Hilbert space, let 
$\tilde{\Gamma}=\{\tilde{\Gamma}(s)|s \in S\}$ be a unitary representation of $S$ in $\mathcal{H}$ [such as Eq.~(\ref{eq:Ttilde})]: There exists a matrix $\tilde{\Gamma}(s)$ for each $s\in S$. The matrices fulfill $\tilde{\Gamma}(s)=\tilde{\Gamma}(s')\tilde{\Gamma}(s'')$ if $s=s'\cdot s''$.
Then, group theory tells us that there exists a basis transformation that brings all representations $\tilde{\Gamma}(s)$ into block-diagonal form
	\begin{eqnarray}
        \tilde{\Gamma}(s)\rightarrow
		\Gamma(s) = \mqty(
  \Gamma_1(s) & 0         & \dots & 0 \\
        0         & \Gamma_2(s) & & \vdots \\
        \vdots    & & \ddots & 0 \\
        0         & \dots & 0 & \Gamma_n(s)
  ).
        \label{eq:GammaBlockDiag}
	\end{eqnarray}
Just as $\tilde{\Gamma}$ and $\Gamma$, the sets $\Gamma_m=\{\Gamma_m(s)|s\in S\}$ ($1\leq m\leq n$) are representations of the group $S$. The transformation is defined such that the $\Gamma_m$ cannot be reduced any further; they are called irreducible representations (irreps). For a given $m$, the matrices $\Gamma_m(s)$ have the same dimensions, they are $1\times 1$, $2\times 2$, or $3\times 3$ matrices. Since the maximal irrep dimension is thus 3, $n$ is of the same order of magnitude as the dimension of $\cal{H}$. On the other hand, we know from group theory that the number of distinct irreps is smaller than or equal to the number of symmetry operations
\footnote{To be precise, there is the relation $|S|=\sum_\mu d_\mu^2$ with the number of symmetry operations $|S|$ and the dimension $d_\mu$ (1, 2, or 3) of the $\mu$th irrep. The sum runs over all irreps.}; 
hence, many of the sets $\Gamma_m$ are identical
\footnote{Our criterion for the basis transformation was the block-diagonality of $\Gamma$. This still permits any two matrix representations, say $\Gamma_m$ and $\Gamma_{m'}$, to be formally different but equivalent; that is, there exists a unitary matrix $U$ such that $\Gamma_m(s)=U^{-1}\Gamma_{m'}(s)U$ for all $s$. In other words, they are in the same equivalence class. It is then straightforward to incorporate $U$ into the basis transformation to make $\Gamma_m$ and $\Gamma_{m'}$ identical.}.

The reader may have noticed that we use the same symbol "$\Gamma$" on the right-hand side of Eq.~(\ref{eq:GammaBlockDiag}) as in Eq.~(\ref{eq:Ts}) for the symmetry transformation of the single-particle wave functions. This choice is more than a mere notational convenience. In fact, provided that the basis sets are suitably chosen, the matrices $\Gamma_m(s)$ and $\Gamma(s)_{\vbk n}^{\vbk'n'}$ are identical if they correspond to the same irrep.

We now re-index the $\Gamma_m$ by $(\mu,i)$ where $\mu$ ($1 \leq \mu \leq r$) is an index for the irrep and $m$ ($1 \leq i \leq n_\mu$) counts the $n_\mu$ subspaces transforming according to the same irrep. Sorting $\Gamma_{\mu,i}$ by ascending  irrep index $\mu$ gives
	\begin{eqnarray}
		\Gamma(s) = \mqty(
        \Gamma_{1,1}(s) & 0         & \dots & 0 \\
        0         & \Gamma_{1,2}(s) & & \vdots \\
        \vdots    & & \ddots & 0 \\
        0         & \dots & 0 & \Gamma_{r,n_r}(s)
  ).
		\label{eq:irrep_matrix}
	\end{eqnarray}
By construction, the $\Gamma_{\mu,m}(s)$ are identical for all $m$, so we may omit the index $m$ in $\Gamma_\mu(s)$.

Now consider a linear operator $H$ on $\mathcal{H}$ that commutes with
all $\Gamma(s)$. Using the decomposition into irreps, we can write $H$ as a matrix consisting of blocks $H^{\mu,i}_{\nu,j}$,
	\begin{eqnarray}
		H = \mqty(
			H^{1,1}_{1,1} & \dots & H^{1,1}_{r,n_r} \\
			\vdots & \ddots & \vdots \\
			H^{r,n_r}_{1,1} & \dots & H^{r,n_r}_{r,n_r}
		).
	\end{eqnarray}
The great orthogonality theorem of group theory  states that $H^{\mu,i}_{\nu,j} = 0$ if $\mu \neq \nu$, i.e., the subspaces of nonequivalent irreps do not couple with each other. 
The operator $H$ is thus block-diagonal with blocks $H_\mu$ defined on
the corresponding subspaces $\mathcal{H}_\mu$ for each irrep $\mu$,
	\begin{eqnarray}
        H = 
        \begin{pmatrix}
        H_{\mu=1} & 0         & \dots & 0 \\
        0         & H_{\mu=2} & & \vdots \\
        \vdots    & & \ddots & 0 \\
        0         & \dots & 0 & H_{\mu=r}
        \end{pmatrix}.
        \label{eq:Hblock}
	\end{eqnarray}
Furthermore, the great orthogonality theorem states that each nonzero block $ H^{\mu,i}_{\mu,j} $ is a multiple of the identity matrix.
Note that $H_\mu$ is potentially sparse but, in general, not a multiple of the identity matrix.
	
\subsubsection{Decomposition into irreps}\label{sec:decomp}
	
To relate these general results of group theory to our present case, the (reducible) representation $\tilde{\Gamma}$ of the symmetry group is given by Eq.\ (\ref{eq:2particle-symrep}), but its decomposition into irreps $\Gamma$ is still unknown. It will not be our goal to determine the irreps explicitly. For our purpose, it will suffice to perform a decomposition of the Hilbert space $\mathcal{H}$ into symmetry-adapted subspaces $\mathcal{H}_\mu$, as this already makes the electron-hole Hamiltonian block-diagonal.

There is a unique canonical decomposition into subspaces. 
Consider a representation $\tilde{\Gamma}$ of the group $S$ on the space $\mathcal{H}$. 
Consider one of the irreps, say with the index $\mu$.
We will now construct the subspace $\mathcal{H}_\mu$.
At first, we define the characters $\chi_\mu(s)$ of the representation $\Gamma_\mu$ by
	\begin{eqnarray}
	    \chi_\mu(s) = \Tr[\Gamma_\mu(s)] \in \mathbb{C}\,.
	\end{eqnarray}
An important property of the characters is that they are orthonormal in the sense
	\begin{eqnarray}
	    \frac{1}{\abs{S}} \sum_s \chi^*_\mu(s) \chi_\nu(s) = \delta_{\mu\nu}\label{eq:LOT}
	\end{eqnarray}
with $|S|$ the number of elements of $S$. Equation (\ref{eq:LOT}) is also called the \emph{little orthogonality theorem}.
In the following, we only need the character tables instead of the complete irreps. Character tables are available for all space groups from the Bilbao crystallography server \cite{Aroyo2006BilbaoCS}, the International Tables for Crystallography \cite{ITA2002}, or similar publications. 
	
With the characters $\chi_\mu(s)$, we can now declare an operator on $\mathcal{H}$,
	\begin{eqnarray}\label{eq:projection}
	    P_\mu = \frac{d_\mu}{\abs{S}} \sum_s \chi^*_\mu(s) \tilde{\Gamma}(s)
	\end{eqnarray}
	where $d_\mu$ is the dimension of 
 the $\mu$th irrep.
	The operator commutes with any matrix $\tilde{\Gamma}(t)$,
	\begin{eqnarray}
	    \tilde{\Gamma}(t)^{-1} P_\mu \tilde{\Gamma}(t)
	    =
	    \frac{d_\mu}{\abs{S}} \sum_s \chi^*_\mu(s) \tilde{\Gamma}(t^{-1} s t)\nonumber
	    \\
	    =
	    \frac{d_\mu}{\abs{S}} \sum_s \chi^*_\mu(t^{-1} s t) \tilde{\Gamma}(t^{-1} s t)
	    = 
	    P_\mu\,.
	\end{eqnarray}
Let us assume that we already know a basis set in which the reducible representations $\tilde{\Gamma}(s)$ block-diagonalize into $\Gamma(s)$ of Eq.\ (\ref{eq:irrep_matrix}), which would amount to replacing $\tilde{\Gamma}\rightarrow\Gamma$ in Eq.\ (\ref{eq:projection}). In this basis, $P_\mu$ is also block-diagonal.
In particular, we have $(P_\mu)^{\nu^\prime i}_{\nu j}=(P_\mu)^{\nu i}_{\nu i}\delta_{\nu\nu'}\delta_{ij}$.
The great orthogonality theorem applies to $P_\mu$ and tells us that each block $(P_\mu)_{\nu,i}^{\nu,i}$ is a multiple of the identity matrix. We can determine the prefactor by taking the trace
	\begin{eqnarray}
	    \Tr[(P_\mu)_{\nu,i}^{\nu,i}] = \frac{d_\mu}{\abs{S}} \sum_s
	    \chi^*_\mu(s) \Tr[\Gamma_{\nu,i}(s)] \nonumber \\
	    = \frac{d_\mu}{\abs{S}}\sum_s \chi^*_\mu(s) \chi_\nu(s) = d_\mu \cdot \delta_{\mu \nu}\,.
	\end{eqnarray}
As a consequence, $(P_\mu)_{\nu,i}^{\nu,i}=0$ if $\mu \neq \nu$, otherwise it is the identity matrix. 
This means that $P_\mu$ is a projection operator that projects an arbitrary function into the subspace $\mathcal{H}_\mu$. 
	
A basis for $\mathcal{H}_\mu$ can thus be obtained by applying $P_\mu$
of Eq.\ (\ref{eq:projection}) to a complete set of vectors. Such a set
is simply given by (1,0,...,0), (0,1,...,0), etc.; each unit vector
corresponds to one of the product basis functions of Eq.~(\ref{eq:2p-product-basis}).
Note that the resulting set of projected functions is highly linearly dependent, in general. Many of the functions will just vanish. Many of them will be identical. 
We use singular value decomposition to obtain an orthonormal basis $\{\vb{v}_{\mu,k}\}_{k \leq n_\mu \cdot d_\mu}$. It may appear computationally costly to apply the projector $P_\mu$ in the product basis space given the large dimensionality of this space. However, the action of $P_\mu$ on a particular product function $\varphi_{\vb{k} u}(\vbr\sigma) \varphi_{\vb{k} o}^*(\vbr'\sigma')$ acts only within the space of the so-called \emph{orbit} associated with that function: the orbit is the set of all product functions obtained by applying the symmetry operations to $\varphi_{\vb{k} u}(\vbr\sigma) \varphi_{\vb{k} o}^*(\vbr'\sigma')$. The orbit consists of all product basis functions, whose constituent occupied and unoccupied states are energy degenerate to $\varphi_{\vb{k} u}$ and $\varphi_{\vb{k} o}$. Each orbit can thus be regarded as an equivalence class that groups together all degenerate product functions. The orbit dimension is much smaller than the product-basis dimension and is at most $|S|d_{\mrm{max}}^2$, where $d_{\mrm{max}}$ is the maximal degree of degeneracy; $d_{\mrm{max}}=3$ is the maximum value among all space groups. Specifically, the number of elements in an orbit does not scale with the size of the $\vbk$ point grid or the number of bands that define the two-particle basis.

By repeating the above procedure for all irrep classes, we obtain orthonormal symmetry-adapted basis sets for all irreps. The subspaces $\mathcal{H}_\mu$ themselves are orthogonal to each other, as well, so that we finally obtain an orthogonal basis transformation matrix
	\begin{eqnarray}
	    U = \mqty(\vb{v}^*_{1,1} \\ \dots \\ \vb{v}^*_{r,n_r d_r})
        \label{eq:transformationU}
	\end{eqnarray}
which decomposes our representation $\tilde{\Gamma}$ into the irreps. The transformation $U^{\dagger}HU$
then yields the block-diagonal form of the electron-hole Hamiltonian of Eq.~(\ref{eq:Hblock}).

The transformation $U^\dagger HU$ can be performed "irrep-wise", $U_\mu^\dagger HU_\mu$. Furthermore, the vectors $\{\vb{v}_{\mu,k}\}$ have maximally  $|S|d_\mrm{max}^2$ nonzero components. This upper bound on the number of nonzero components does not increase with the system size or the $\mathbf{k}$ grid. The transformation matrix $U$ is therefore sparse.

\begin{figure}
  \centering
  \includegraphics[width=0.45\textwidth]{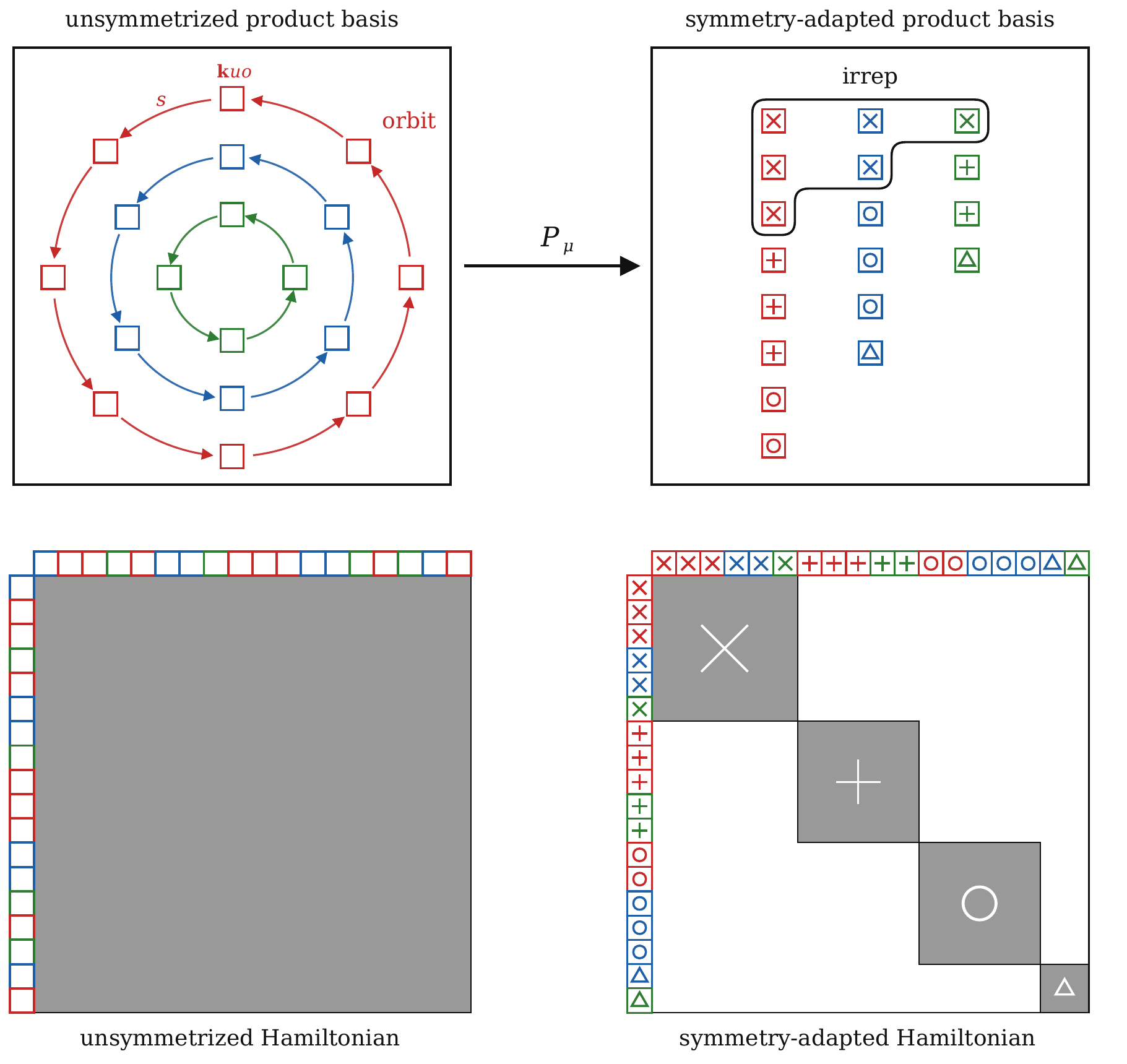}
  \caption{\label{fig:schematic}Schematic of the construction of the symmetry-adapted product basis from the unsymmetrized one via the projectors $P_\mu$. The resulting block-diagonal form of the symmetry-adapted electron-hole Hamiltonian is shown below next to the dense unsymmetrized Hamiltonian. See text for details.}
\end{figure}

Figure \ref{fig:schematic} shows a schematic of the construction of the symmetry-adapted basis. Each square in the upper left panel stands for a basis function of Eq.~\ref{eq:2p-product-basis} of the original unsymmetrized product basis. One of the squares is labeled with the corresponding indices $\{\vbk uo\}$. The product basis functions (initially in arbitrary order) can be grouped into orbits by application of the symmetry transformations $s$ of the space group. The colors red, blue, and green denote different orbits. Each orbit is then projected into each of the irrep subspaces using the projector of Eq.~\ref{eq:projection}. The symbols "$\times$", "+", "$\circ$", and "$\triangle$" in the upper right panel denote different irreps. If the projector for irrep "$\times$" is applied to the product functions of the red orbit, we would get the three symmetry-adapted product basis functions marked with a red "$\times$"; the other projections would vanish. By projecting all orbits with every irrep we successively build up the symmetry-adapted product basis, eventually representing a unitary transform of the original unsymmetrized basis. All symmetry-adapted product functions that transform according to one of the irreps form the basis of that irrep (see, e.g., the line enclosing all functions marked with "$\times$"). The lower left panel represents the dense electron-hole Hamiltonian in the original unsymmetrized product basis. The upper row and left column of squares represent the product basis functions in no particular order. The lower right panel shows the block-diagonal form of the Hamiltonian represented in the symmetry-adapted product basis ordered irrep-wise, one block per irrep. The matrix elements in the white regions are all zero.

\subsection{Symmetry-reduced BSE Hamiltonian}
\label{subsec:symmetry-reduction}

\subsubsection{Optical absorption spectrum}
 

	
	


Diagonalizing the $r$ blocks $U_\mu^\dagger H U_\mu$ with a parallel solver gives the full set of eigensolutions of $H$, categorized into irreps. 
Assuming that all $r$ blocks have the same size, one would estimate a
maximum theoretical computational speedup of $r^2$ with respect to the diagonalization of the full matrix $H$: $r$ blocks, each diagonalization costs $1/r^3$ relative to the full matrix. However, as discussed below, not all irreps are relevant for optical absorption. In fact, in the present examples, we find that only one of the irreps contributes to the absorption spectrum, while all others do not. The estimated maximal theoretical speedup for the diagonalization is therefore as large as $r^3$. 
 
    
 The decomposition naturally classifies the excitonic excitations into
 irreps, thereby determining their symmetry properties without the
 need for an \emph{a posteriori} analysis \cite{Nalabothula2025}. Each
 of the obtained eigenfunctions belongs to a single irrep, which
 guarantees that their symmetry properties are fully preserved and no averaging step is needed in the visualization. In the present systems, all optically active excitons belong to the same irrep. The respective irrep and the corresponding number of excitonic excitations can be identified prior to the actual calculation.
	
Other approaches that circumvent the diagonalization of the electron-hole Hamiltonian, for example, the Lanczos-Haydock method \cite{Haydock80,Gruening2011} or the time-evolution approach \cite{schmidt2003,hahn_quasiparticle_2005} also benefit from the sparse block-diagonal form of $H$, as they involve a large number of matrix operations. In both methods,
the final absorption spectrum is simply given by the sum of the contributions calculated for each contributing block separately. Here, too, only one irrep may need to be taken into account.


Equation (\ref{eq:macro_diel}) together with Eq.~(\ref{eq:reducible_polar}) yields the macroscopic dielectric function $\varepsilon_\mathrm{M}(\omega)$, whose imaginary part gives the optical absorption spectrum. The eigensolutions enter Eq.~(\ref{eq:reducible_polar}) with the eigenvalues $\Omega_\lambda$ and eigenvectors $A^\lambda_{\vbk uo}$. To be more precise, only those eigenvectors contribute to the absorption spectrum for which Eq.~(\ref{eq:oscillatorA}) is nonzero. Let us write Eq.~(\ref{eq:oscillatorA}) as
$\sum_{\vbk uo}A^\lambda_{\vbk uo}B_{\vbk uo}$ and, expressed in the symmetry-adapted basis, as $\sum_{k}A^\lambda_{\mu k}B_{\mu k}$. 

For the systems studied in this work, we find that $B_{\mu k}=0$ ($1\leq k\leq n_\mu$) for all irreps $\mu$ except one. This is not a coincidence but rather a consequence of group theory and the underlying symmetry group. Since the operator $-i(\hat{\vbq}\nabla)$ transforms as the vector $\hat{\vbq}$, we consider the irrep decomposition of the vector representation $\hat{\mathbf{x}}$, $\hat{\mathbf{y}}$, and $\hat{\mathbf{z}}$ with respect to the present symmetry group, resulting in maximally three irreps (because the vector representation is three-dimensional). If all vectors transform according to the same irrep $\mu$, then $B_{\mu' k}=0$ for all $\mu'\neq\mu$, which is a consequence of the great orthogonality theorem of group theory. This situation occurs in Si and LiF. The same conclusion holds when the vector representation decomposes into two or three irreps but $\hat{\vbq}$ belongs to only one of them, as in bulk MoS$_2$. Only if $\hat{\vbq}$ has nonzero projections onto two or three irreps do those irreps contribute to optical absorption (but never more than three irreps). The latter only happens in the case of low-symmetry systems and/or if the polarization vector $\vbq$ is chosen along a direction that is not aligned with a symmetry axis.

\subsubsection{Exciton energies and loss spectrum}

So far, we have discussed the poles of the optical absorption spectrum
or macroscopic dielectric function $\epsilon_\mathrm{M}(\omega)$, which are given by the eigenvalues of the electron-hole Hamiltonian Eq.\,(\ref{eq:BSE-eigval}) corresponding to transversal excitons. Another set of eigenvalues, the longitudinal excitons, forms the pole structure of the \emph{inverse} macroscopic dielectric function $\epsilon^{-1}_\mathrm{M}(\omega)$. Its imaginary part is proportional to the electron-energy loss spectrum \cite{Rost2023}. A similar derivation as the one leading to Eq.\,(\ref{eq:BSE-eigval}), gives rise to an eigenvalue equation for $\epsilon^{-1}_\mathrm{M}(\omega)$ identical in structure to Eq.\,(\ref{eq:BSE-eigval}) but with the important difference that the modified Coulomb potential $\overline{v}$ should be replaced by the real, unmodified Coulomb potential $v$.
    
As a consequence, a term arising from the Coulomb divergence eliminated for $\overline{v}$ has to be added to Eq.\,(\ref{eq:BSE-eigval}) to get the Hamiltonian for the electron-energy loss spectrum. This term can be extracted from Eq.\,(\ref{eq:w_interaction2}) expressed in a plane-wave basis instead of the mixed basis $\{M_{\vbq I}(\vbr)\}$. The Coulomb matrix is then diagonal, and the double sum over $IJ$ becomes a single sum over $\vbG$. The desired term is the one for $\vbG=\mathbf{0}$, for which the Coulomb matrix element diverges as $v_\vbq\stackrel{\vbq\rightarrow\mathbf{0}}{\sim}4\pi/q^2$. Luckily, the other two terms are linear in $q$, 
\begin{equation}
\frac{1}{\sqrt{V_\mathrm{uc}}}\lim_{q\rightarrow 0}
\langle e^{i\vbq\vbr}\varphi_{\vbk n}| \varphi_{\vbk+\vbq m}\rangle=
\frac{1}{\sqrt{V_\mathrm{uc}}}
\vbq 
\frac{\langle \varphi_{\vbk n} | -i\nabla | \varphi_{\vbk m} \rangle}{\epsilon^0_{\vbk m}-\epsilon^0_{\vbk n}},
\end{equation}
so that the $\vbG=\mathbf{0}$ term is finite
\begin{eqnarray}
\nonumber
	v^\mathrm{div}_{\vbk u,\vbk' o';\vbk o,\vbk' u'} &=& \frac{4\pi}{N_\vbk V}
        \frac{\langle \varphi_{\vbk u} | -i(\hat{\vbq}\nabla) | \varphi_{\vbk o} \rangle}
        {\epsilon^0_{\vbk u}-\epsilon^0_{\vbk o}}\\
        &&\times
        \frac{\langle \varphi_{\vbk' o'} | -i(\hat{\vbq}\nabla) | \varphi_{\vbk' u'} \rangle}
        {\epsilon^0_{\vbk' u'}-\epsilon^0_{\vbk' o'}}\nonumber
        \\
        &=&\frac{4\pi}{N_k V}B^*_{\mathbf{k}uo}B_{\mathbf{k}'u'o'}\label{epsinv_div}
\end{eqnarray}       
    with $\hat{\vbq}=\vbq/q$, the polarization direction of the incoming beam of light. On the right-hand side, we encounter again the quantity defined in Eq.~(\ref{eq:oscillator}).

Equation (\ref{epsinv_div}), the term to be added to Eq.\,(\ref{eq:BSE-eigval}), will, in general, affect all matrix elements. It seems that this gives rise to yet another eigenvalue equation of the same dimensions, which would double the cost of solving the BSE (if we want the eigenvalue spectrum of both the optical absorption spectrum and the electron-energy loss spectrum). 
Fortunately, transforming the resulting modified electron-hole Hamiltonian $H'$ to the eigenbasis of the original (unmodified) Hamiltonian $H$ recovers a classical problem of linear algebra with a well-known solution. The resulting secular equation becomes the condition $\operatorname{Re}[\varepsilon_\mrm{M}(\Omega_\xi')]=0$, giving the eigenvalues $\Omega_\xi'$ of $H'$ as the poles of the inverse macroscopic dielectric function $\varepsilon^{-1}_\mrm{M}(\omega)$.

Before we discuss the secular equation, we first have a closer look at the modified eigenvalue problem. The transformation to the eigenspace of $H$ yields $v^\mrm{div}_{\lambda\lambda'}=4\pi/(N_k V)B^*_\lambda B_{\lambda'}$.
If one of the $B_\lambda$, say for the $\lambda$th eigenvector, vanishes, then the $\lambda$th row and $\lambda$th column of $H'$ will be identical to those of $H$ (with all elements zero except the diagonal one). Thus, the $\lambda$th eigensolution of $H$ will also be an eigensolution of $H'$ (with zero coupling to light, i.e., a dark exciton). Such eigenvectors can thus be eliminated from the eigenvector basis.
For the same reason, only the eigensolutions of the irrep(s) relevant for the optical absorption spectrum (see previous paragraph) need to be taken into account.
Finally, given a set of $n$ degenerate eigensolutions of $H$, we can always make a unitary transformation among them such that only one of the $B_\lambda$ is nonzero and the others vanish. The corresponding eigensolutions $\{\Omega_\lambda,A^\lambda\}$ with $B_\lambda=0$ are then eigensolutions to $H'$, too, and can be eliminated from the eigenvalue problem of $H'$. 
We thus end up with an eigenvalue problem, in which all eigenvalues can be assumed to be nondegenerate.

Applying the matrix determinant lemma to the characteristic equation of $H'$ in the eigenspace of $H$, $\mathrm{det}[H-\Omega'I+4\pi/(N_k V)\mathbf{B}^*\mathbf{B}^\mathrm{T}]=0$ with the identity matrix $I$, leads to the secular equation
\begin{equation}
1-\frac{4\pi}{N_k V}\sum_\lambda \frac{|B_\lambda|^2}{\Omega'-\Omega_\lambda}=0,\label{eq:secular}
\end{equation}
which corresponds to the condition $\operatorname{Re}[\varepsilon_\mrm{M}(\Omega'_\xi)]=0$ for the eigenvalues $\Omega'_\xi$ of the modified Hamiltonian.
The function on the left-hand side has poles at the unperturbed
eigenvalues $\Omega_\lambda$. Between two consecutive eigenvalues,
$\Omega_\lambda$ and $\Omega_{\lambda+1}$, the function is strictly monotonically decreasing so that there is a solution $\Omega'_\xi$ between each pair of unperturbed eigensolutions. There is an additional solution above the highest unperturbed eigenvalue. The lowest $\Omega_\xi'$ (with a nonzero oscillator strength) is thus strictly higher than the lowest unperturbed eigenvalue $\Omega_\lambda$. Numerically, the nodes can be found efficiently, e.g., by interval nesting or the Newton method. We use the LAPACK routine DLASD4, which, after suitable preparation of the input data, can be used for the purpose of solving the secular equation (\ref{eq:secular}). 

The corresponding eigenvectors $\mathbf{A}^{\prime\xi}$ fulfill
\begin{equation}
\left(H+\frac{4\pi}{N_k V}\mathbf{B}^*\mathbf{B}^\mathrm{T}\right)\mathbf{A}^{\prime \xi}=\Omega_\xi'\mathbf{A}^{\prime\xi},
\end{equation}
from which we can derive the closed form
\begin{equation}
A^{\prime\xi}_\lambda=\frac{B^*_\lambda}{\Omega_\lambda-\Omega'_\xi}\left(-\frac{4\pi}{N_k V}\mathbf{B}^\mrm{T}\mathbf{A}^{\prime\xi}\right)\label{eq:eigenvec_xi}
\end{equation}
for its components in the eigenbasis of $H$. The bracket on the right-hand side is unknown but nonzero by construction. Since eigenvectors are defined only up to a factor, this factor is irrelevant, and the eigenvector is fully defined by Eq.~(\ref{eq:eigenvec_xi}). Multiplying with $A_\lambda^{\prime \xi}$ and summing over $\lambda$ identifies Eq.~(\ref{eq:eigenvec_xi}) simply as the condition $\operatorname{Re}[\varepsilon_\mrm{M}^{-1}(\Omega_\lambda)]=0$, the companion condition to $\operatorname{Re}[\varepsilon_\mrm{M}(\Omega_\xi')]=0$: The poles of the macroscopic dielectric function are the nodes of its inverse $\varepsilon^{-1}_\mrm{M}(\omega).$
Calculating the eigenvalue spectrum from the secular Eq.~(\ref{eq:secular}) and the eigenvectors from Eq.~(\ref{eq:eigenvec_xi}) takes negligible computation time.


\subsection{Parallel storage}

Because of the quadratic scaling of the number of product basis functions with respect to the system size and its cubic scaling with respect to the number of $\vbk$ points, it is essential to parallelize the code in order to treat large systems and/or achieve high accuracy. In particular, the large size of the electron-hole Hamiltonian requires the matrix to be stored in parallel across multiple compute nodes, since the limited memory of a single node is quickly exhausted.

We use a two-step process. In the first step, the Hamiltonian $H$ is calculated in the standard two-particle product basis. When $H$ is complete, it is transformed to the symmetry-adapted basis in a second step. Both matrices are stored in distributed memory, which requires a suitable data distribution strategy using MPI as described in the following.

As we use the Tamm-Dancoff approximation, the Hamiltonian matrix $H$ is Hermitian, and we can afford to just store its upper triangle in the first step. The storage of this upper triangle is evenly distributed across the compute nodes. Each node keeps its part in shared memory (MPI 3.1), i.e., shared among the processes on that node. All processes of the node can access this memory by load/store operations as if it were an array allocated by the process itself. 

The memory demand for storing the electron-hole Hamiltonian with dimension $N$ on $M$ nodes is thus $N(N+1)/(2M)$ real or complex floating-point numbers per node. This amounts to 305~GB per node in the most expensive LiF calculation with a 36$\times$36$\times$36 $\vbk$ mesh and $N=1\,982\,862$ running on 48 nodes. In principle, keeping only the relevant irrep block in memory instead of the full Hamiltonian would further reduce the memory demand. However, we were unable to find an efficient algorithm to perform the required transformation to the symmetry-adapted basis on the fly for nonconsecutive matrix elements. Therefore, we perform the transformation after constructing the full Hamiltonian in one go immediately before diagonalization.

The calculation of the matrix elements is parallelized over the
processes. All processes independently calculate matrix elements with
a minimum of communication between the processes (and without
calculating any element twice). However, because of to the usage of symmetries in the calculation of the matrix elements, see Sec.~\ref{subsec:sym_mat}, the matrix elements calculated by a given process are in general not contiguous in memory. It is therefore not possible for the processes to calculate only the matrix elements that are stored in the memory of the local node. All processes eventually need to send their values to all nodes (except their own).

Here, we use the following strategy: 
A given process writes the matrix elements with fast store operations either to the (node-)local part of the matrix or to a buffer array. (Each process has $M-1$ buffer arrays, where $M$ is the number of nodes.)
There is an outer loop in the calculation of the BSE direct and exchange matrix elements, namely over the $\vbq$ points (of the irreducible wedge) in the former [corresponding to the contribution of $W(\vbq)$] and over the processes in the latter. At the end of each loop iteration, several send-receive (MPI Sendrecv) operations are executed to transmit buffer data to the corresponding remote nodes, where a local process receives it and writes it to the local portion of the matrix stored on that node.
In this way, the communication always transfers contiguous memory arrays (the buffer arrays) and the (noncontiguous) writing to the matrix is performed by fast store operations.
The communication overhead is well below the computation time of the matrix elements. Transferring the data to the nodes requires auxiliary send and receive arrays containing at most 
$(M-1)N(N+1)|S|/(2MPN_\vbk)$ and $(M-1)N(N+1)/(2MP^2)$ floating-point numbers per MPI process for the direct and exchange terms, respectively, where $P$ denotes the number of MPI processes. 
For the LiF calculation, the auxiliary arrays  require about 629~MB and 100~MB per node, respectively.

In the second step, $H$ is transformed to the symmetry-adapted basis
using the (sparse) transformation matrix (\ref{eq:transformationU}.
The result of the transformation is stored in block-cyclic form as
required by the parallel linear-algebra libraries ScaLAPACK
\cite{Choi96} or ELPA \cite{Auckenthaler11,Marek14}. Specifically, the right-multiplication of the $U$ matrix is calculated in a round-robin fashion giving, in each iteration and process, a very thin matrix with $b$ columns where $b$ is the chosen block size, typically 64, 96, or 128. The left-multiplication of $U^\dagger$ is then done by the same process in chunks of $b$ rows so that, in each step, the result is a complete $b$$\times$$b$ block, which is then sent to the process that owns that block (according to the block-cyclic distribution). Before calculating the next block, each process receives its own blocks from other processes. In this way, the transformed Hamiltonian is immediately stored in block-cyclic form and can be passed to the parallel linear algebra library. In this step, the communication overhead can be significant for large matrix sizes (see Sec.~\ref{sec:results}).

During the transformation, both the full and the symmetry-reduced electron-hole Hamiltonians must coexist in memory. As required by the block-cyclic layout for ScaLAPACK or ELPA, the matrix to be diagonalized must reside fully in memory, not just its upper or lower triangle. The additional memory demand is then $N'^2/M$ per node where $N'$ is the dimension of the symmetry-reduced Hamiltonian (e.g., $N' \approx N/5$ for silicon). In the LiF example, this requires 22~GB of additional memory per node.

\section{Results and Discussion} \label{sec:results}
	

To verify the code and to demonstrate the speedup achieved, we show
results for bulk silicon, lithium fluoride, and molybdenum disulfide. For these materials, there are numerous BSE studies in the literature for comparison. For each material, we show the convergence of the lowest peaks of the optical absorption ($E_\mrm{opt}$) and the electron-energy loss spectra ($E_\mrm{b}$) relative to the conduction band minimum. These values correspond to the transversal and longitudinal exciton binding energies, respectively.

\begin{figure}
  \centering
  \includegraphics[angle=-90,width=0.45\textwidth]{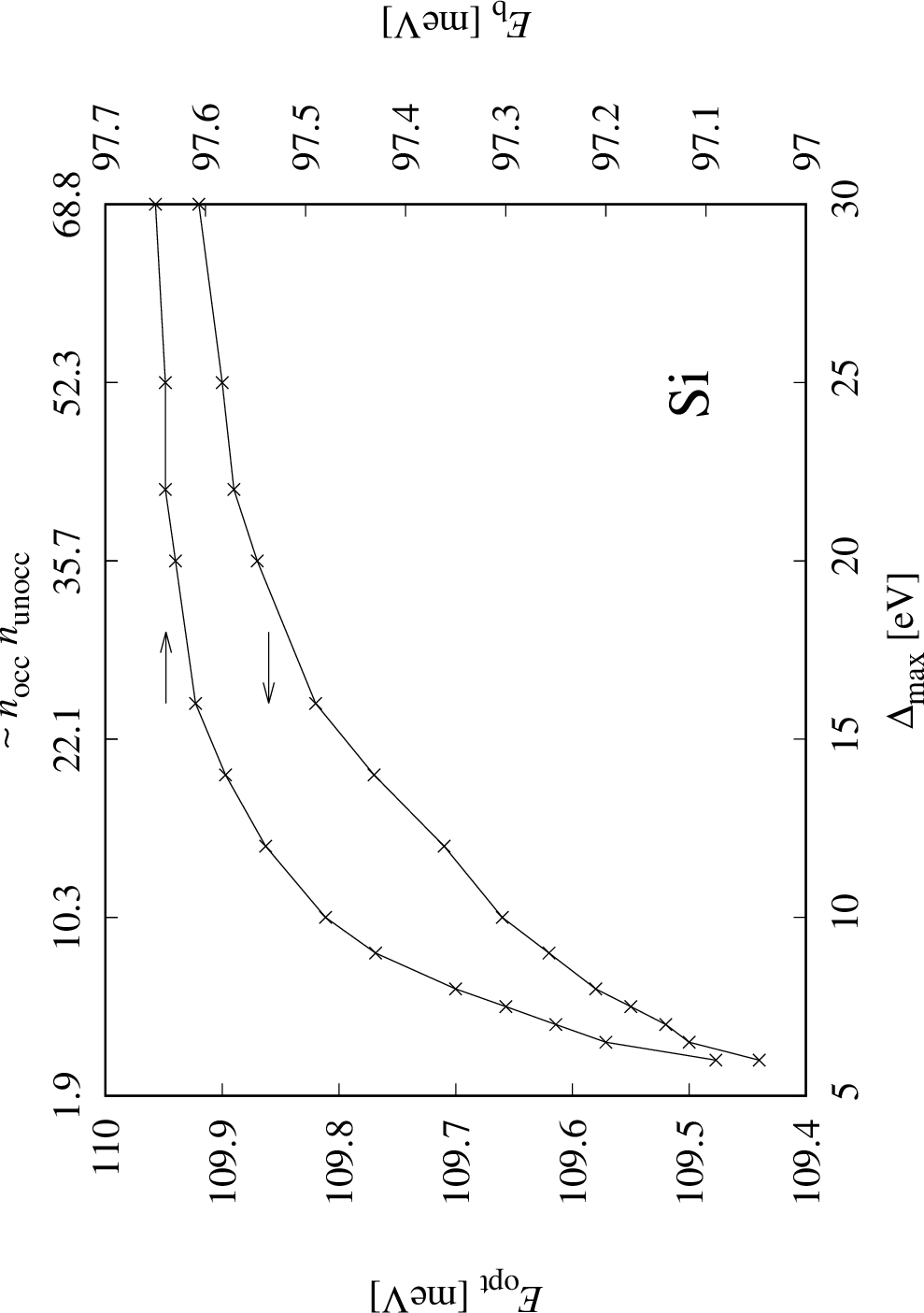}
  \caption{\label{fig:ecutconv_Si}Convergence of the exciton binding energies $E_\mrm{opt}$ and $E_\mrm{b}$ with respect to the product-basis cutoff parameter $\Delta_\mrm{max}$. The upper $x$ axis shows the average product $n_\mrm{occ}n_\mrm{unocc}$ (see text).}
\end{figure}

The noninteracting reference system is obtained from a standard Kohn-Sham DFT calculation with the Perdew-Burke-Ernzerhof functional \cite{Perdew96}. Since the focus here is on the numerical method rather than the results 
and to facilitate reproducibility of the results, we apply a simple scissors shift to open the band gap  to (approximately) the $GW$ gap, instead of performing a  $GW$ calculation beforehand. 

The scissor shift simply shifts the entire optical spectrum rigidly up or down in energy. It can thus be used to align the absorption onset of the theoretical spectrum with the experimental one. The exciton binding energy, on the other hand, is an energy difference and is therefore unaffected by such a shift. Specifically, a positive scissor shift $\Delta$ increases all diagonal elements of the electron-hole Hamiltonian and, consequently, all eigenvalues by $\Delta$. At the same time, it increases the band gap by the same amount $\Delta$. The exciton binding energy, defined as the difference between these two quantities, therefore remains unchanged. As is common in BSE calculations, we employ the unscissored DFT Green function to evaluate the screened interaction. Consequently, the screening is likewise unaffected by the scissor operator. It should be noted that the symmetry decomposition of the Hamiltonian depends only on the irreducible representations of the space group and on the single-particle wave functions forming the electron-hole product basis. It is therefore independent of the scissor operator or any other quasiparticle treatment.

We first turn to bulk silicon. The experimental lattice constant of 5.43\,\AA{} is used. The ground-state calculation is carried out with a \kgrid{16} $\vbk$-point grid. Further computational parameters are the reciprocal cutoff radius of 4.0\,bohr$^{-1}$ and an $l$ cutoff of 8. The MT radius is 2.15\,bohr. The Kohn-Sham DFT direct ($\Gamma-\Gamma$), $\Gamma-\mrm{X}$, and fundamental band gaps of 2.56\,eV, 0.71\,eV, and 0.57\,eV, respectively, underestimate the experimental gaps. To align the direct band gap approximately with the experimental one ($3.3~\mrm{eV}$ \cite{Philipp60}), we apply a scissor operator in the calculation of the electron-hole Hamiltonian that uniformly shifts the conduction bands up by 0.75\,eV. A total of 100 bands are included in the calculation of the screened interaction. Both the bare and the screened interaction ($v$ and $W$) are expanded in the mixed basis with a reciprocal cutoff radius of 3.0\,bohr$^{-1}$ and an $l$ cutoff of 4. We use the formalism of the random-phase approximation for $W$ without resorting to a plasmon-pole model or similar simplifications. 

Instead of specifying the numbers of occupied and unoccupied states
from which the product basis is constructed, we employ an energy
cutoff $\Delta_\mrm{max}$ and include all products of Eq.\,\ref{eq:2p-product-basis} into the basis for which $\epsilon^0_{\vbk u}-\epsilon^0_{\vbk o}<\Delta_\mrm{max}$. The advantage is that there is only one parameter to converge (instead of two). Another advantage is that degenerate states are always either fully included in the product basis or excluded. This avoids cutting degenerate subspaces, which would otherwise result in symmetry breaking. 

Figure\ \ref{fig:ecutconv_Si} shows the convergence of $E_\mrm{opt}$ and $E_\mrm{b}$ with respect to $\Delta_\mrm{max}$ calculated with an 8$\times$8$\times$8 $\vbk$-point sampling. The upper $x$ axis further indicates the quantity $n_\mrm{occ}n_\mrm{unocc}$ averaged over the $\vbk$ grid. To be more precise, this quantity corresponds to the number of product basis functions of Eq.\ (\ref{eq:2p-product-basis}) divided by the number of $\vbk$ points ($8^3=512$ in this case).
It provides a link to the usual product-basis specification in the literature, which is expressed in terms of $n_\mrm{occ}$ and $n_\mrm{unocc}$ (together with the $\vbk$ grid).
For example, if $n_\mrm{occ}=n_\mrm{unocc}=2$, then $n_\mrm{occ}n_\mrm{unocc}=4$ should be compared to the values given on the upper $x$ axis of the diagram. We observe a very fast convergence with respect to $\Delta_\mrm{max}$. Both $y$ axes show a tiny energy range of less than 1\ meV. We use $\Delta_\mrm{max}=7.5~\mrm{eV}$ for our calculations in the following, which corresponds to  $n_\mrm{occ}n_\mrm{unocc}\approx 5.6$.

The number of $\vbk$ points is an important convergence parameter for
BSE calculations. It is known that exciton binding energies converge slowly with the $\vbk$ grid. On the other hand, the number of $\vbk$ points strongly impacts the computational cost. The diagonalization of the BSE electron-hole Hamiltonian scales as $n^9$ for an \kgrid{n} $\vbk$-point sampling. Furthermore, the construction of the screened interaction scales as $n^6$. Since DFT calculations usually converge quickly with the $\vbk$ grid, we store the self-consistent effective potential obtained with the \kgrid{16} set on hard disc and use it for all BSE calculations. Each calculation starts with the generation of the Kohn-Sham eigensolutions $\{\varphi_{\vbk n},\epsilon_{\vbk n}\}$ on the respective $\vbk$ grid. The screened interaction is calculated and expanded in the mixed basis, then projected onto the BSE product basis according to Eq.~\ref{eq:w_interaction} yielding the direct term of Eq.~\ref{eq:BSE-eigval}. Likewise, the exchange term of Eq.~\ref{eq:BSE-eigval} is calculated with the corresponding mixed-basis expansion of the bare interaction. The so-constructed electron-hole Hamiltonian is transformed to the symmetry-reduced Hamiltonian according to Sec.~\ref{sec:decomp}, which is then diagonalized using a standard parallel diagonalizer (ScaLAPACK or ELPA).

    \begin{table}
        \begin{tabular}{l|r|r|r|r}
            \multicolumn{5}{c}{Si}\\
            \hline
			$\vbk$ grid & full dim & block dim & $E_\mathrm{opt}$ [meV] & $E_\mrm{b}$ [meV] \\
			\hline
			\kgrid{6} & 1 284 & 264 & 137.6 & 115.3 \\
			\kgrid{8} & 2 868 & 561  & 109.6 & 97.3 \\
			\kgrid{12} & 9 978 & 1 917  & 73.2 & 67.9 \\
			\kgrid{16} & 22 902 & 4 380 & 54.4 & 51.4 \\
		    \kgrid{20} & 45 344 & 8 637 & 43.7 & 41.5 \\
		    \kgrid{24} & 78 990 & 14 988 & 36.8 & 35.1 \\
		    \kgrid{30} & 153 246 & 28 998 & 30.4 & 28.8 \\
		    \kgrid{36} & 264 194 & 49 905 & 26.5 & 24.9 \\
		    \kgrid{42} & 418 688 & 79 002 & 24.3 & 22.3 \\
		    \kgrid{50} & 707 756 & 133 401 & 22.9 & 20.4 \\
            \kgrid{60} & 1 220 864 & 229 887 & 22.5 & 19.5 \\
            \hline
            Exp. \cite{Green2013} & --- & --- & 15\hphantom{.0} & --- \\
		\end{tabular}

		\caption{\label{tab:Si}
			BSE calculations of silicon for different $\mathbf{k}$-point sets. We report the number of rows (columns) of the full electron-hole Hamiltonian (full dim) as well as the number of rows (columns) of the symmetry-reduced Hamiltonian block (block dim). The last two columns contain the exciton binding energies appearing as the lowest peaks in the absorption spectrum ($E_\mrm{opt}$) and the electron-energy loss spectrum ($E_\mrm{b}$) as well as the experimental value. 
		}
	\end{table}

Table \ref{tab:Si} shows the $\vbk$-point convergence of the transverse and longitudinal exciton binding energies relevant for the absorption ($E_\mrm{opt}$) and EEL spectrum ($E_\mrm{b}$). The matrix dimensions of the full Hamiltonian and the symmetry-reduced Hamiltonian are shown in the second and third column, respectively. The latter amounts to about 20\,\% of the former. Given that numerical diagonalization scales cubically with the matrix dimension, this would result in a theoretical acceleration factor of 125 for this step, which is, however, in practice reduced somewhat by the quadratic and linear scaling terms. Both $E_\mrm{opt}$ and $E_\mrm{b}$ converge slowly with respect to the $\vbk$ grid. 


\begin{figure}
  \centering
  \includegraphics[angle=-90,width=0.45\textwidth]{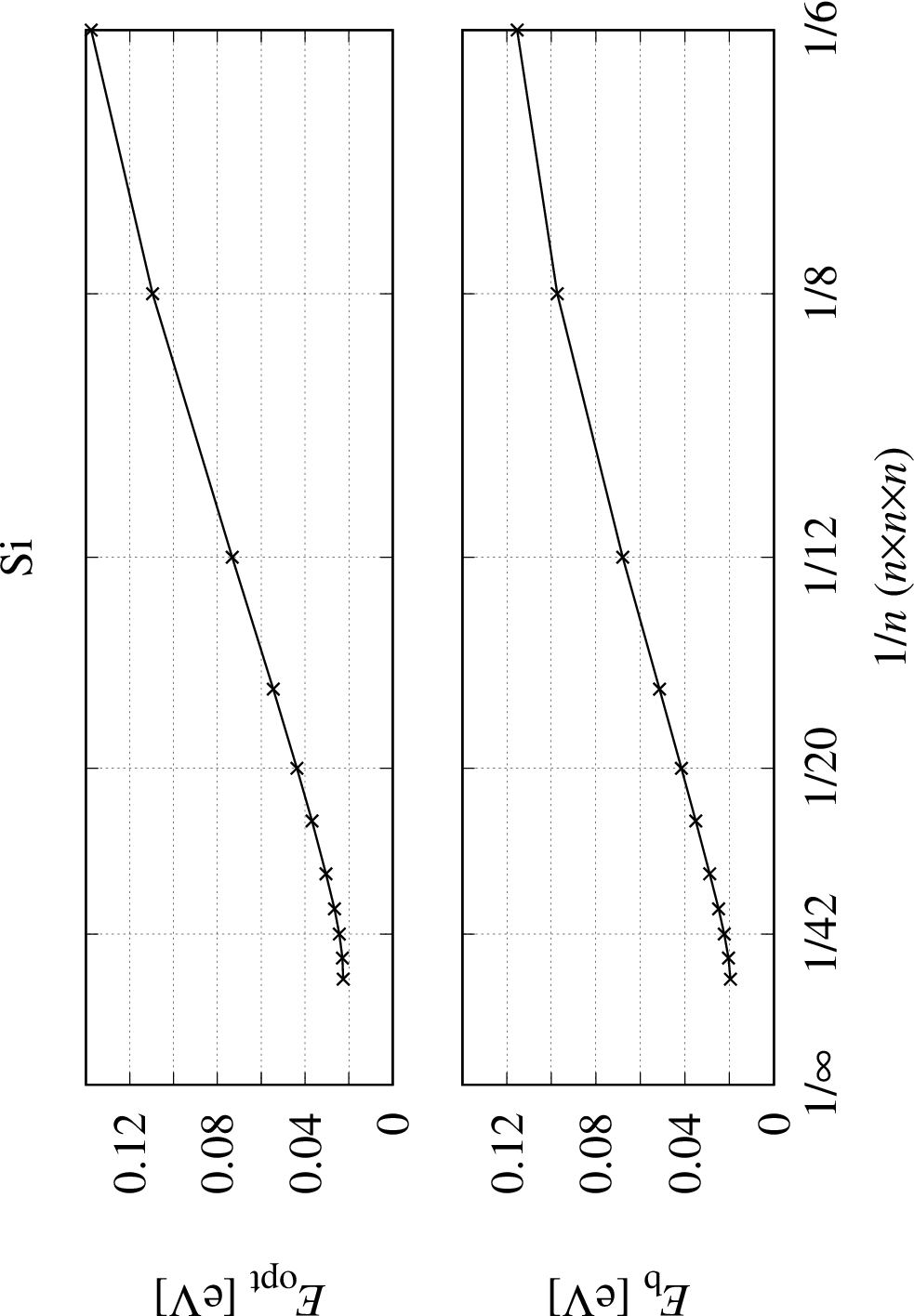}
  \caption{\label{fig:kconv_Si}Convergence of the exciton binding energies $E_\mrm{opt}$ and $E_\mrm{b}$ with respect to the $\vbk$ grid \kgrid{n} as a function of $1/n$ for Si.}
\end{figure}

Figure~\ref{fig:kconv_Si} shows the behavior of $E_\mrm{opt}$ and $E_\mrm{b}$ with respect to the $\vbk$ grid \kgrid{n} as a function of $1/n$. The energies do not exhibit a clear analytic behavior for increasingly dense $\vbk$ meshes that would allow for an extrapolation to the infinite-mesh limit. We therefore consider the values at \kgrid{60} to be our best estimates. In fact, the curves level off at this point indicating that the values are already close to the infinite-mesh limit.

The first peak $E_\mrm{opt}=22.5~\mrm{meV}$ of the optical absorption
spectrum still overestimates the experimental value of 15\,meV
\cite{Green2013} but is closer to experiment than previous ab initio
BSE binding energies \cite{Sun2020}. The result of 22.5\,meV is close to the recently published exciton binding energy of 24\,meV calculated based on a nonuniform Brillouin-zone sampling scheme \cite{Alvertis2023}. Given the slow $\mathbf{k}$ convergence of the exciton binding energies, the better agreement may result from the use of a denser $\mathbf{k}$ mesh. Nevertheless, the present data are insufficient to draw this conclusion with certainty.

\begin{figure}
  \centering
  \includegraphics[angle=-90,width=0.45\textwidth]{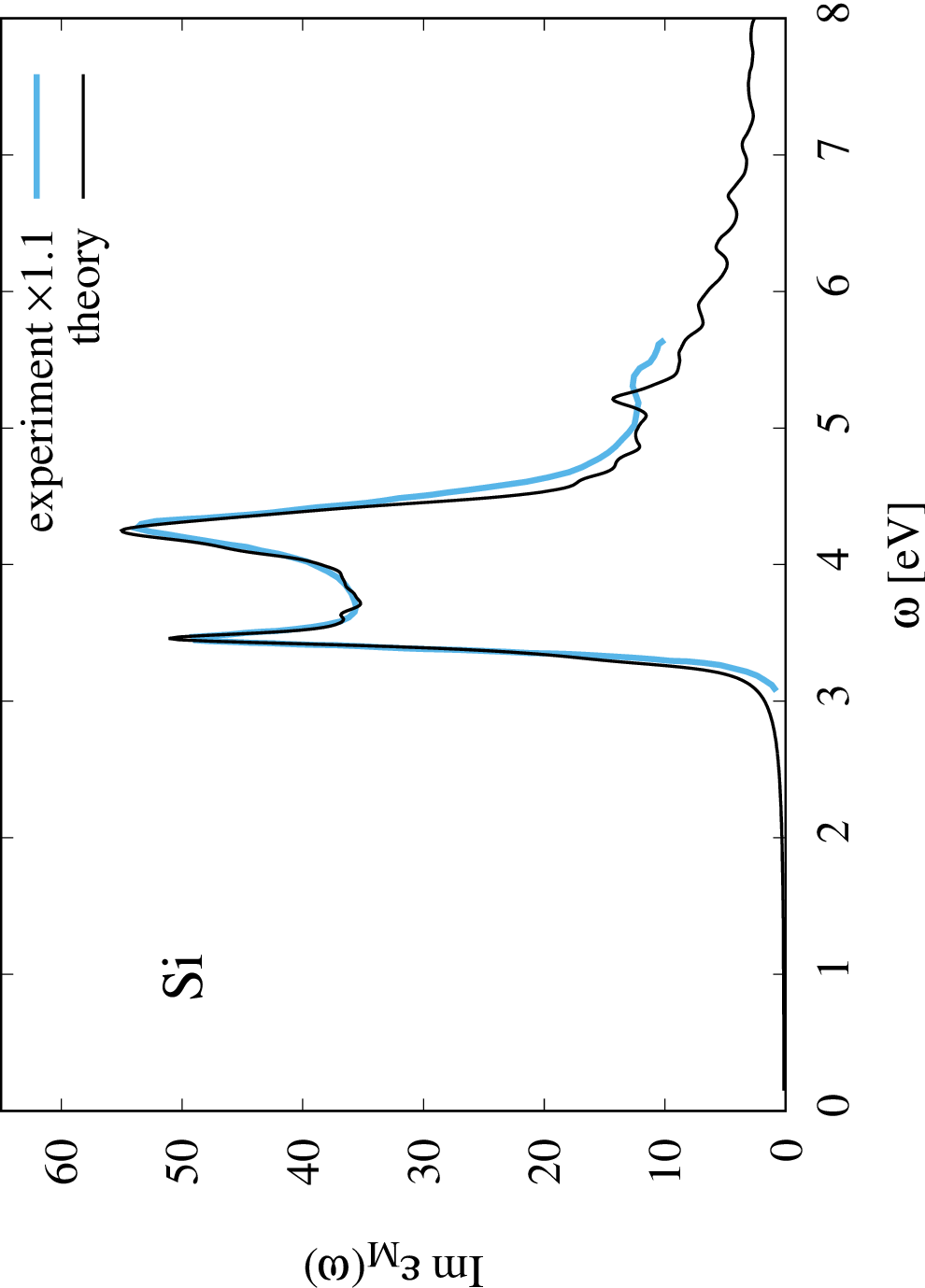}
  \caption{\label{fig:exp_Si} Comparison of the theoretical
  (\kgrid{60}) and experimental absorption spectra \cite{Lautenschlager87} for Si.  The experimental spectrum was vertically scaled by a factor of 1.1.
  }
\end{figure}

Figure~\ref{fig:exp_Si} presents a comparison of the theoretical
absorption spectrum $\mrm{Im}\varepsilon_\mrm{M}(\omega)$ calculated
with a \kgrid{60} $\vbk$ mesh to the experimental absorption spectrum
from Ref.~\onlinecite{Lautenschlager87}. The experimental spectrum was
vertically scaled by a factor of 1.1. The shapes of the two spectra
agree very well. Both spectra also exhibit a small shoulder at roughly
the same energy around $5.2~\mrm{eV}$. It should be noted that the
fact that the peaks align so well horizontally is because of the choice of the scissor operator. For example, increasing the scissor operator from 0.75~eV to 0.80~eV would shift the theoretical spectrum by 50~meV to the right.

\begin{table}
\begin{tabular}{l|r|r}
  \multicolumn{3}{c}{Computational timings}\\
  \hline
  & with sym. & w/o sym. \\
  \hline
  irreps & 8.8 & --- \\ \hline
  direct term & &\\
  - construction & 159.1 & 7924.4 \\
  - communication & 2.0 & 59.7 \\ \hline
  exchange term & &\\
  - construction & 18.6 & 22.9 \\
  - communication & 0.8 & 0.6 \\ \hline
  BSE solution & &\\  
  - block-cyclic & 4.1 & 12.2 \\
  - diagonalization & 4.5 & 342.2
\end{tabular}
\caption{\label{tab:timings}
  Computational timings (in seconds) for the different stages of the
  calculation for bulk silicon using a 20$\times$20$\times$20 $\vbk$
  mesh. The calculations are performed in parallel on a single compute
  node with 128 physical cores, using one MPI process per core. (Left) Accelerated calculation employing the proposed symmetry-assisted evaluation of the matrix elements (Sec.~\ref{subsec:sym_screen}) and the symmetry-reduced BSE Hamiltonian (Sec.~\ref{subsec:symmetry-reduction}). (Right) Standard calculation without the proposed accelerations.
  The wall-clock times are consistently about 1\% larger.
  The benchmark is included for illustrative purposes; it is not intended as a systematic scaling study.
  (Compiler: Intel Fortran Compiler 2021.10.0; libraries: IntelMPI 2021.10, ELPA 2023.05.001; CPU: AMD EPYC 7742)
}
\end{table}

In Table \ref{tab:timings}, we compare the computational timings with and without the proposed symmetry-based accelerations for several stages of the calculation: (1) "irrep": calculation of the irreps  $\Gamma(s)_{\vbk n}^{\vbk' n'}$ and definition of the symmetry-adapted product basis; (2) "construction": calculation of the direct and exchange terms; (3) "communication": transfer of the calculated matrix elements to the distributed-memory Hamiltonian array using MPI send-receive operations; (4) "block-cyclic": transformation of the Hamiltonian matrix to the symmetry-adapted basis while simultaneously distributing the block-cyclic portion to each process (again requiring MPI communication); (5) "diagonalization": diagonalization of the relevant block of the electron-hole Hamiltonian.
We want to emphasize that "calculation without symmetry" refers to the restriction of $\vbk$ and $\vbk'$ to IBZ and EIBZ($\vbq$) [or EIBZ($\vbk$)], respectively, in the calculation of the Hamiltonian matrix elements (Sec.~\ref{subsec:sym_screen}), and the symmetry-reduced Hamiltonian matrix (Sec.~\ref{sec:decomp}). In all calculations, however, $\vbq$ in $W(\vbq)$ is restricted to the IBZ, as is standard practice in BSE codes.

The construction of the irreps and the symmetry-adapted basis takes less than nine seconds. This step scales favorably with system size, so its relative impact on the total computation time will be even smaller for larger systems. The most expensive step is the calculation of the direct term, which takes about 2 m 30 s. Here, the speedup factor with respect to a calculation without symmetry is about 50. The communication required to update the Hamiltonian matrix with the direct term also shows a large difference between the two calculations: 2 s versus 60 s. This is surprising since the dimension of the Hamiltonian matrix is identical in the two cases and corresponds to the full product basis of Eq.~\eqref{eq:2p-product-basis}. (The transformation to the symmetry-adapted basis is carried out only immediately before the diagonalization.) The reason for the different timings is that the total work is divided into $N$ packets, where $N$ is the number of irreducible $\vbk$ points in the former (256) and all $\vbk$ points in the latter (8000). The larger number of packets increases the effects of load imbalance and computational overhead, resulting in the observed difference in timings.

The exchange term requires much less computation than the direct term, even though the same number of matrix elements is calculated. Consequently, exploiting symmetry has only a minor effect on the computational cost, which in this case is largely determined by memory bandwidth rather than arithmetic operations. Since the amount of data transferred is essentially the same in both calculations, the corresponding timings are also similar (18.6 s and 22.9 s). For larger systems, where the computational workload becomes more significant, exploiting symmetry is expected to yield a larger speedup. In contrast to the direct term, transferring the exchange-term matrix elements to the distributed-memory array takes nearly the same time in the two calculations. This is because the number of packets in this case is chosen to be equal to the number of MPI processes, which is identical in the two cases. 

Finally, the diagonalization of the electron-hole Hamiltonian benefits
greatly from the symmetry-reduced Hamiltonian block. The dimension of
the matrix block is about a factor of five smaller than in the
calculation without symmetry. However, the observed speedup factor is
76, rather than $5^3=125$. The reason is that the cost of
diagonalization is not purely cubic in the matrix dimension but also
contains quadratic and linear contributions, not least because of MPI communication. 
As the matrix size increases, the cubic term becomes increasingly dominant. For small and intermediate matrix sizes, however, the lower-order terms remain significant, resulting in a speedup factor smaller than 125. For larger systems, the diagonalization becomes increasingly expensive relative to the other stages of the calculation.

Right before the diagonalization, the Hamiltonian matrix is transformed to the symmetry-adapted basis and distributed in a block-cyclic manner across the MPI processes. In the present case, this procedure takes nearly as much time as the diagonalization itself. We have found this step to be relatively costly. One possible reason is that only the upper triangle of the Hermitian Hamiltonian is stored. While this saves us about 50\% of memory, it inevitably leads to load imbalance during the transformation to the symmetry-adapted basis. We intend to address this in a later version of the code.


The second test material is lithium fluoride. We employ the
experimental lattice constant of 4.026\,\AA{}. The DFT calculation is
carried out with a reciprocal cutoff radius of 4.4\,bohr$^{-1}$ and $l$ cutoff values of 9 and 7 for Li and F, respectively. The $1s$ core state of lithium is treated as a semicore state with a local orbital. We also add $s$ and $p$ local orbitals to the fluorine atom at high energies, to improve the description of high-lying states. The Kohn-Sham band gap is 9.05\,eV. To account for the underestimation of the band gap in DFT, we will employ a scissors shift as in the case of Si. However, as the experimental value of the direct band gap is not known accurately, we will determine the scissor operator later by aligning the theoretical and experimental absorption spectra.
The band summations for the polarization function comprise all available bands (160-190 bands). The mixed basis is constructed with a reciprocal cutoff radius of 3.3\,bohr$^{-1}$ and $l$ cutoff values of 5 and 4 for Li and F, respectively. 

\begin{figure}
  \centering
  \includegraphics[angle=-90,width=0.45\textwidth]{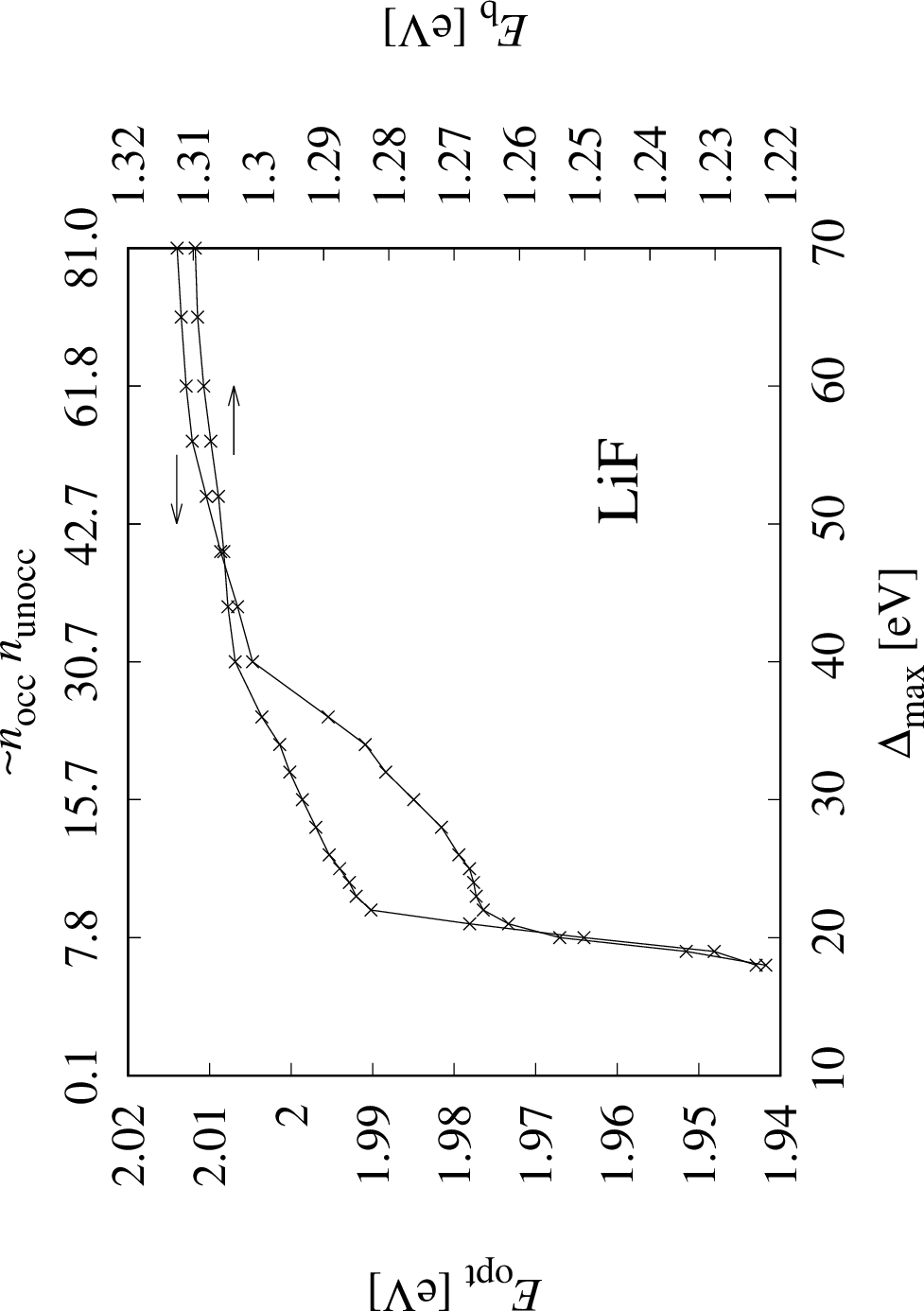}
  \caption{\label{fig:ecutconv_LiF}Same as Fig.~\ref{fig:ecutconv_Si} for LiF.}
\end{figure}

Figure~\ref{fig:ecutconv_LiF} shows the convergence of $E_\mrm{opt}$ and $E_\mrm{b}$ with respect to $\Delta_\mrm{max}$ for LiF. 
Because of the large band gap, the screening is weak giving rise to a strong interaction between the electron and the hole. The exciton binding energy is therefore much larger than in the case of silicon.
The convergence is slower than for Si, as is evident from the markedly different energy scales: 100~meV for LiF vs.~less than 1 meV for Si. As a consequence, we have to choose a much larger $\Delta_\mrm{max}$ parameter than in the Si case. We employ $\Delta_\mrm{max}=50$~eV, which corresponds to $n_\mrm{occ}n_\mrm{unocc}\approx 42.7\approx 6\times 7$.

    \begin{table}
		\begin{tabular}{l|r|r|r|r}
            \multicolumn{5}{c}{LiF}\\
            \hline
			$\vbk$ grid & full dim & block dim & $E_\mathrm{opt}$ [eV] & $E_b$ [eV] \\
			\hline
			\kgrid{6} & 9 200 & 1 860 & 1 983 & 1 350 \\
			\kgrid{8} & 21 838 & 4 338  & 2 009  & 1 306 \\
			\kgrid{12} & 73 426 & 14 313  & 2 066  & 1 344 \\
			\kgrid{16} & 174 224 & 33 636 & 2 099 & 1 375 \\
		      \kgrid{20} & 339 984 & 65 253 & 2 120 & 1 395 \\
		      \kgrid{24} & 587 680 & 112 362 & 2 133 & 1 408 \\
		      \kgrid{30} & 1 147 662 & 218 580 & 2 147 & 1 422 \\
		      \kgrid{36} & 1 982 862 & 376 662 & 2 157 & 1 432 \\
            $\infty$ & $\infty$ & $\infty$ & 2 202 & 1 475 \\
            \hline
            Exp. \cite{Piacentini75}& --- & --- & 2 090 & --- \\
            Exp. \cite{Piacentini76} & --- & --- & 1 900 & ---
		\end{tabular}
  
		\caption{\label{tab:LiF}
			Same as Table \ref{tab:Si} for lithium fluoride. The row marked with $\infty$ reports the extrapolated exciton binding energies.
		}
	\end{table}

The $\vbk$-point convergence of $E_\mrm{opt}$ and $E_\mrm{b}$ is
presented in Table~\ref{tab:LiF}. The densest $\vbk$-point set is
\kgrid{36}. The number of product basis functions for this
$\vbk$-point set reaches nearly 2 million. The symmetry-reduced
Hamiltonian, however, has a much smaller dimension of only 0.38
million. Diagonalizing a matrix of this size is feasible on a modern
supercomputer. Overall, we see a slower convergence than in the case
of Si, but because of the larger energies, a faster relative convergence. The difference between $E_\mrm{opt}$ and $E_\mrm{b}$ is quite large and quickly converges to about 725~meV.

\begin{figure}
  \centering
  \includegraphics[angle=-90,width=0.45\textwidth]{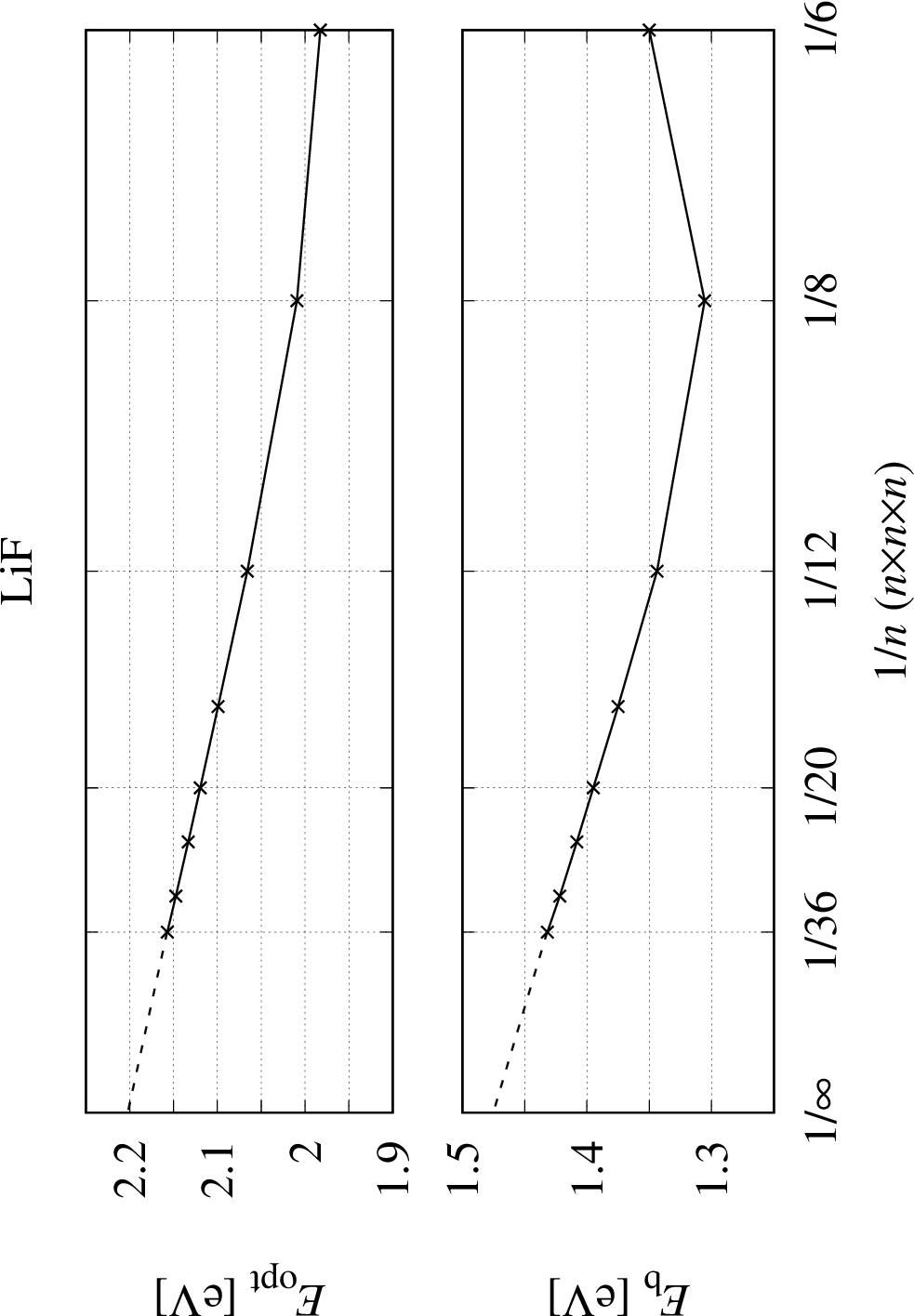}
  \caption{\label{fig:kconv_LiF}Same as Fig.~\ref{fig:kconv_Si} for LiF. The linear extrapolation to the infinite-mesh limit is indicated by dashed lines.}
\end{figure}

Figure~\ref{fig:kconv_LiF} presents a graphical representation of the $\vbk$-mesh convergence. In contrast to Si, linear extrapolation to the infinite-mesh limit is possible and yields the values given in the row marked "$\infty$" in Table~\ref{tab:LiF}. The extrapolated exciton binding energy  $E_\mrm{opt}$ is relatively close to the experimental value of 2.09~eV \cite{Piacentini75}. It should be noted, however, that $E_\mrm{opt}$ is not known accurately in experiment \cite{Piacentini76}. The experimental value of 2.09~eV might thus be a bit misleading in terms of its apparent accuracy. Theoretical studies \cite{rohlfing_electron-hole_2000,Wang03,Deilmann18} have consistently reported smaller $E_\mrm{opt}$ between 1.5 and 1.6~eV. However, a recent BSE study reported 2.05~eV using a \kgrid{16} grid, close to our value of 2.10~eV with this grid. The approximate $0.6$~eV spread among the theoretical values calls for a comparative study of the various implementations in order to examine the numerical differences in detail.
Rohlfing and Louie \cite{rohlfing_electron-hole_2000} report 0.5~eV
for the difference $E_\mrm{opt}-E_\mrm{b}$ in reasonable agreement with our value.

\begin{figure}
  \centering
  \includegraphics[angle=-90,width=0.45\textwidth]{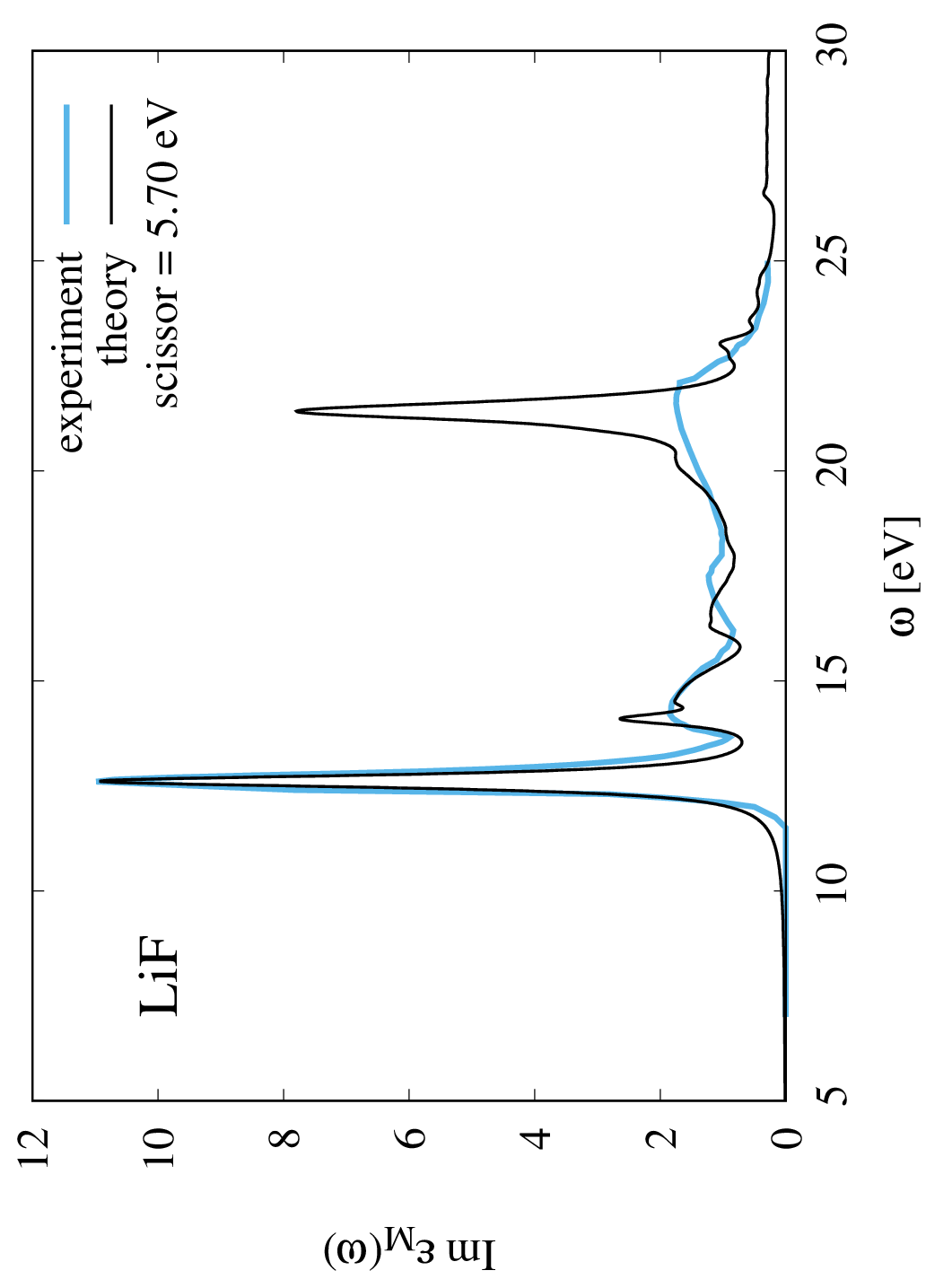}
  \caption{\label{fig:exp_LiF}Comparison of the theoretical (\kgrid{36}) and experimental absorption spectra \cite{Roessler67} for LiF.}
\end{figure}

Figure~\ref{fig:exp_LiF} presents a comparison between the theoretical
absorption spectrum calculated with a \kgrid{36} grid and the
experimental absorption spectrum of Ref.~\onlinecite{Roessler67}, in
which the two components of the refractive index $N(\omega)=n(\omega)+i\kappa(\omega)=\sqrt{\varepsilon_\mrm{M}(\omega)}$ are plotted and tabulated. We thus plotted $\mrm{Im}N^2(\omega)$ as the experimental absorption spectrum. To align the two spectra, we employed a scissor operator of 5.7~eV and a Lorentzian broadening parameter of 0.18~eV for the theoretical spectrum. The sharp $1s$ exciton peak at 12.6~eV is well reproduced. There is a broader peak around 14.5~eV, with a similar shape in the two spectra. A shallow feature at 17~eV appears at a somewhat lower energy in the theoretical spectrum. A sharp peak in the theoretical spectrum at 21.4~eV appears as a broad feature in the experimental spectrum. Both spectra drop off at around 22.2~eV in a similar manner.

As a final test system, we consider bulk MoS$_2$, which crystallizes
in a hexagonal lattice with layers stacked along the $z$ axis. We employ the lattice constants $a=3.16$~\AA{} and $c=12.3$~\AA{} and an  internal parameter of 0.623. The DFT calculation is carried out with the PBE functional on a 12$\times$12$\times$3 $\vbk$-point mesh. The MT radii for Mo and S are 2.41~bohr and 1.99~bohr, respectively. The reciprocal cutoff radius is set to 4.00~bohr$^{-1}$. The cutoff values for the $l$ quantum number are 9 for Mo and 8 for S. The Mo $4s$ and $4p$ states are described by local orbitals. Spin-orbit coupling is included consistently throughout the DFT and BSE calculations. The direct and indirect DFT band gaps are 1.64~eV and 0.87~eV, respectively, which underestimate the experimental gaps by a few 100~meV. We will later employ a scissor operator as in the case of LiF. The band summations for the polarization function 
include 200 bands, ensuring that the exciton binding energies are converged to within 1 meV with respect to this parameter.
For the mixed basis, we employ a reciprocal cutoff radius of 3.0~bohr$^{-1}$ and $l$ cutoff values of 5 and 4 for Mo and S, respectively.

\begin{figure}
  \centering
  \includegraphics[angle=-90,width=0.45\textwidth]{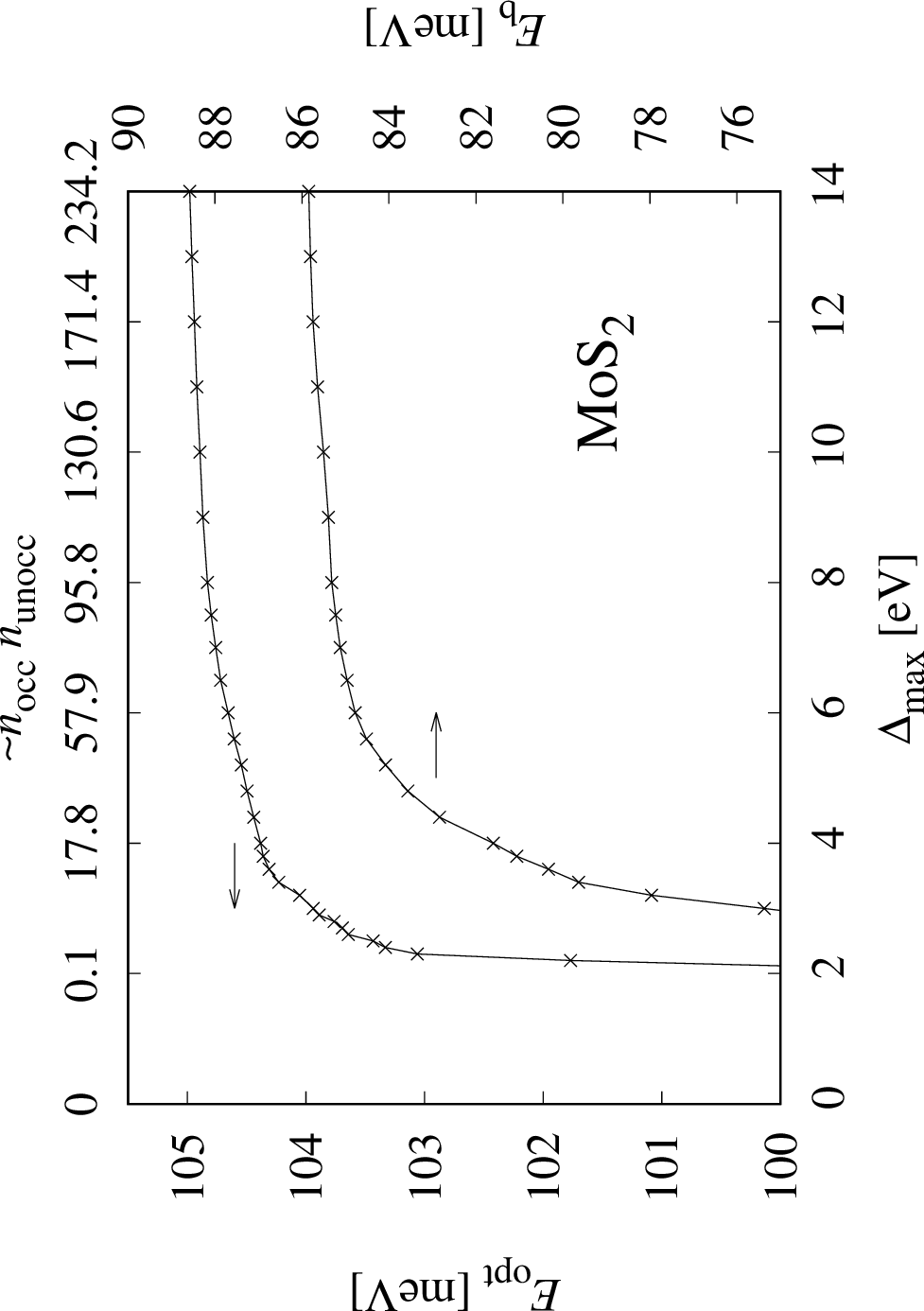}
  \caption{\label{fig:ecutconv_MoS2}Same as Fig.~\ref{fig:ecutconv_Si} for bulk MoS$_2$.}
\end{figure}

Figure \ref{fig:ecutconv_MoS2} shows the convergence of the exciton binding energies $E_\mrm{opt}$ and $E_\mrm{b}$ as a function of the product-basis cutoff energy $\Delta_\mrm{max}$. The convergence is faster than in LiF but not as fast as in the case of Si. For $E_\mrm{opt}$, convergence to within 1~meV is reached at around $\Delta_\mrm{max}=4~$eV, whereas $E_\mrm{b}$ converges a bit more slowly. Since we are mostly interested in the optical spectra, we adopt $\Delta_\mrm{max}=4~$eV for the BSE calculations, which corresponds to $n_\mrm{occ}n_\mrm{unocc}=17.8$; i.e., typically four occupied and four unoccupied bands are included in the product basis.

    \begin{table}
		\begin{tabular}{l|r|r|r|r}
            \multicolumn{5}{c}{MoS$_2$}\\
            \hline
			$\vbk$ grid & full dim & block dim & $E_\mathrm{opt}$ [meV] & $E_b$ [meV] \\
			\hline
			6$\times$6$\times$3 & 7 892 & 1 330 & 290 & 256 \\
			12$\times$12$\times$3 & 30 980 & 5 178  & 134  & 118 \\
			18$\times$18$\times$3 & 68 996 & 11 514 & 95 & 84 \\
		    24$\times$24$\times$3 & 123 644 & 20 622 & 81 & 70 \\
		    30$\times$30$\times$3 & 193 028 & 32 186 & 76 & 65 \\
		    42$\times$42$\times$3 & 378 284 & 63 062 & 76 & 64 \\
            \hline
            Exp. \cite{Beal72} & --- & --- & 50 & --- \\
            Exp. \cite{Fortin74} & --- & --- & 87 & --- \\
            Exp. \cite{Saigal16}& --- & --- & 84 & --- \\

		\end{tabular}
  
		\caption{\label{tab:MoS2}
            Same as Table~\ref{tab:Si} for bulk MoS$_2$.
		}
	\end{table}

\begin{figure}
  \centering
  \includegraphics[angle=-90,width=0.45\textwidth]{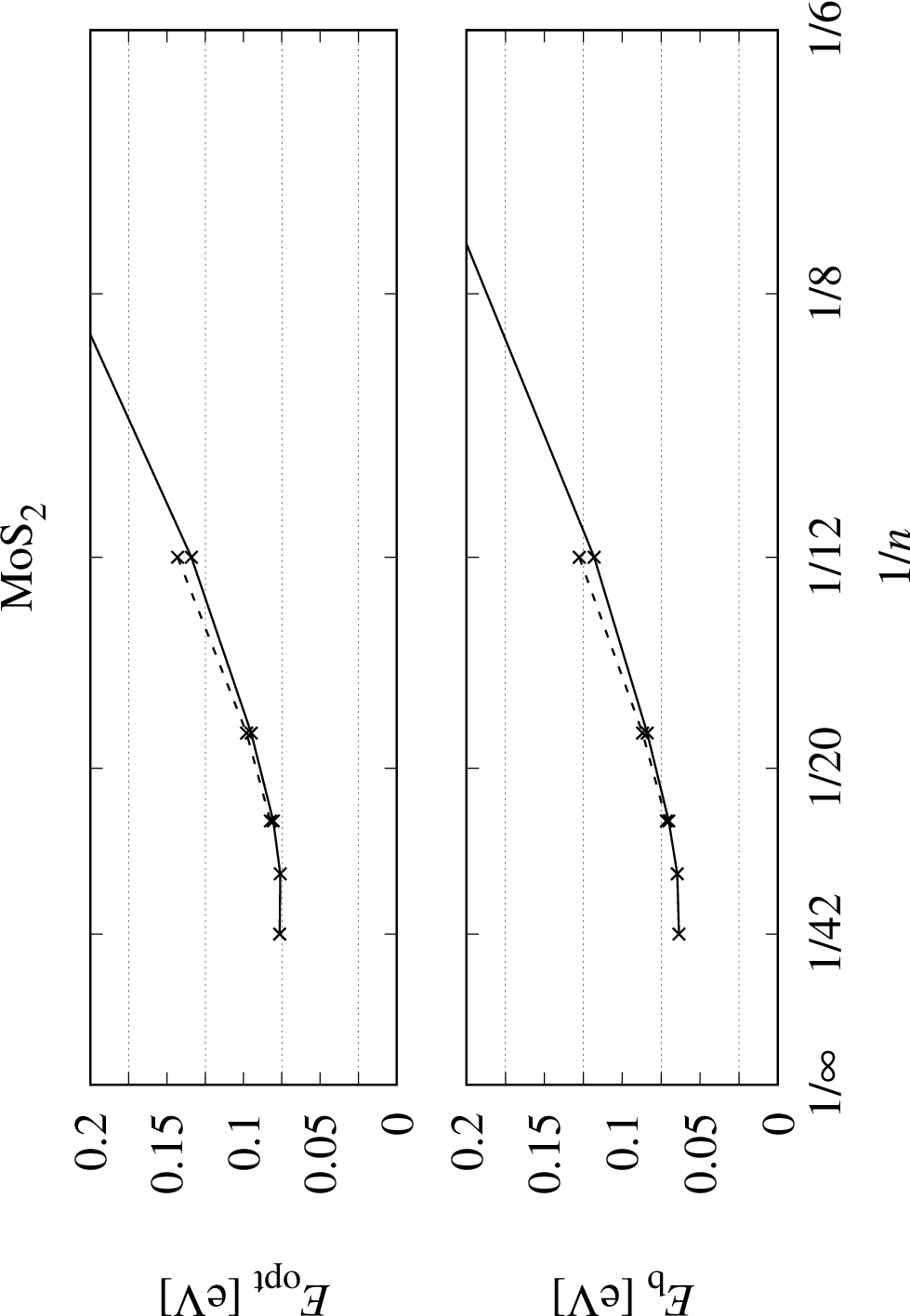}
  \caption{\label{fig:kconv_MoS2}Same as Fig.~\ref{fig:kconv_Si} for bulk MoS$_2$. The solid (dashed) line refers to the $\vbk$-point sampling $n$$\times$$n$$\times$3 ($n$$\times$$n$$\times$5).}
\end{figure}

Table \ref{tab:MoS2} shows the convergence of the exciton binding
energies $E_\mrm{opt}$ and $E_\mrm{b}$ with respect to the
$\vbk$-point sampling. As in the case of Si, the convergence is slow,
in particular for the in-plane $\vbk$-point set, whereas the out-of-plane $\vbk$ convergence is faster (see Fig.~\ref{fig:kconv_MoS2}). The resulting exciton binding energy of 76~meV is in good agreement with the experimental values of 50~meV \cite{Beal72}, 87~meV \cite{Fortin74} and 84~meV \cite{Saigal16}, the latter being from a recent study. Our theoretical value differs by less than 10~meV from this measurement. 
Only a limited number of BSE studies exist for bulk MoS$_2$. A commonly cited reference (Ref.~\onlinecite{Komsa12}) reports an exciton binding energy of 130 meV, significantly higher than our value. This discrepancy likely arises from the coarse 12$\times$12$\times$3 $\vbk$-point set used in that study. Indeed, our calculated value of 134~meV for the same mesh is in very good agreement with theirs.
Another study \cite{molina-sanchez_effect_2013} thoroughly
investigates the $\vbk$-point convergence of MoS$_2$. The authors
focus on the monolayer and bilayer MoS$_2$ and also show absorption spectra for the bulk system, but they do not report an explicit value for the exciton binding energy.

The second and third columns of Table \ref{tab:MoS2} show the dimensions of the full electron-hole Hamiltonian and the symmetry-reduced effective block. The latter is about one-sixth of the dimension of the full Hamiltonian, corresponding to a theoretical 216-fold speedup for the leading cubic scaling term of the diagonalization.

Figure~\ref{fig:kconv_MoS2} shows a graphical representation of the exciton binding energies as a function of the $\vbk$-point sampling $n$$\times$$n$$\times$3. As in the case of Si, the convergence curves flatten at the densest $\vbk$-point mesh of 42$\times$42$\times$3, indicating that the corresponding value should be close to the infinite-mesh limit. We also included $E_\mrm{opt}$ and $E_\mrm{b}$ values calculated with the denser out-of-plane sampling $n$$\times$$n$$\times$5, shown as dashed lines. The two curves are very close to each other, in particular, towards denser in-plane $\vbk$-point meshes, demonstrating that the out-of-plane sampling with three momenta is sufficient.

\begin{figure}
  \centering
  \includegraphics[angle=-90,width=0.45\textwidth]{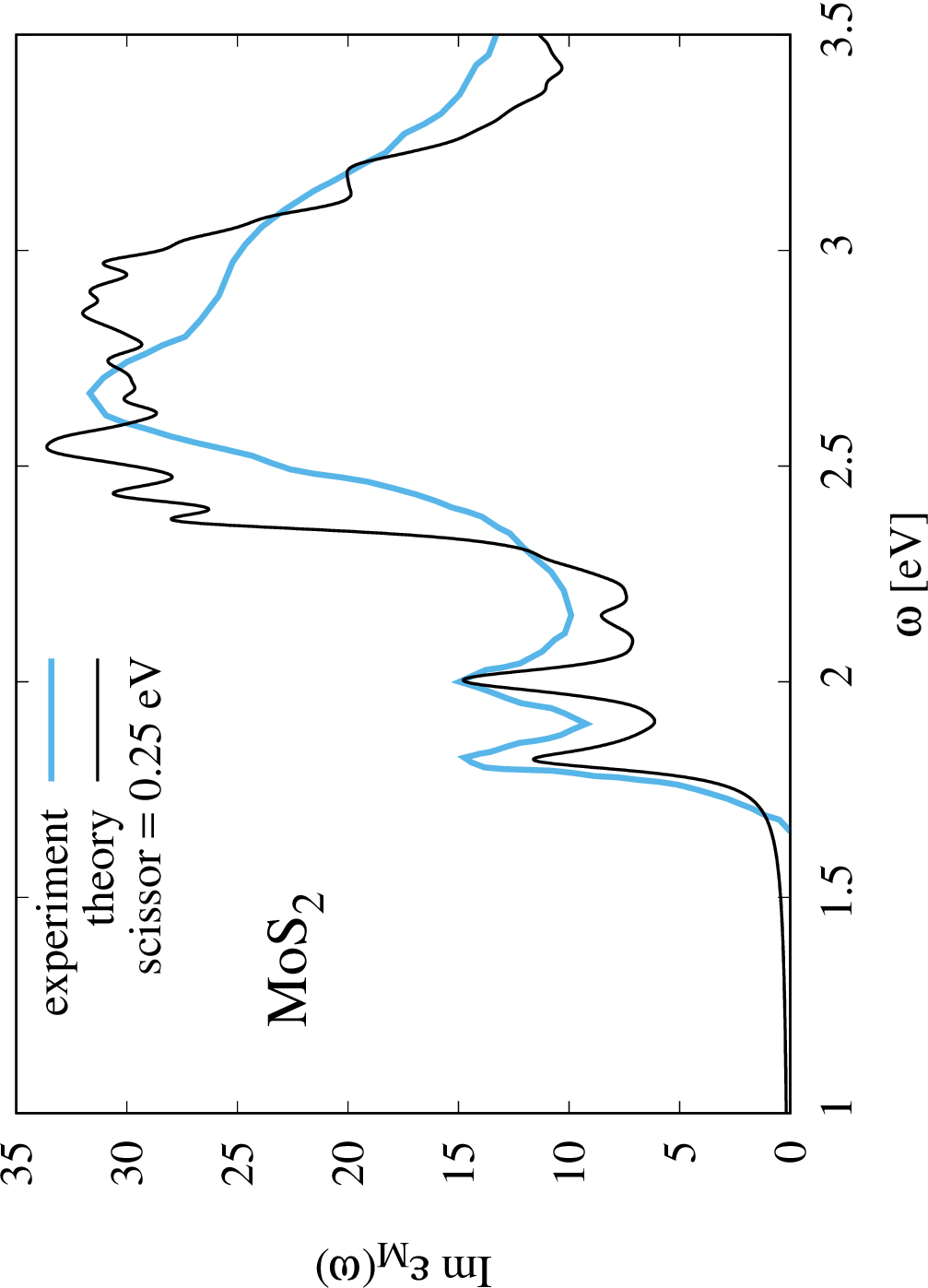}
  \caption{\label{fig:exp_MoS2}Comparison of the theoretical
  (42$\times$42$\times$3) and experimental absorption spectra \cite{Beal79} for bulk MoS$_2$.}
\end{figure}

Finally, Fig.~\ref{fig:exp_MoS2} compares the theoretical optical
absorption spectrum, computed using a 42$\times$42$\times$3
$\mathbf{k}$-point mesh, with the experimental spectrum reported in Ref.~\onlinecite{Beal79}. The theoretical spectrum is scissor-shifted by 0.25~eV. The typical double-peak structure at the onset of the spectrum consisting of excitons A and B is well reproduced. The peak of exciton A is slightly smaller than that of exciton B in the theoretical spectrum, whereas in the experiment the two peaks have approximately the same height. Between 2.4~eV and 3.1~eV, both spectra exhibit a broad feature with a similar intensity of more than twice that of excitons A and B. In the experiment, this feature appears somewhat narrower than in the theoretical spectrum. Additionally, a small peak is visible in the theoretical spectrum between this broad feature and exciton B, which is absent in the experimental data. However, an earlier experimental study (Ref.~\onlinecite{Beal72}) does report a faint feature in this region.
	
\section{\label{sec:conclusions}Summary}

In this paper, we described our implementation of the Bethe-Salpeter equation (BSE) approach for calculating optical absorption and electron-energy loss spectra within the all-electron full-potential linearized augmented-plane-wave (FLAPW) method. To represent the bare and screened Coulomb interaction potentials, we employ the mixed basis, which is derived from the LAPW basis and adapted to represent products of wave functions.

It was demonstrated how crystal symmetries can be exploited to accelerate  the construction and diagonalization of the electron-hole Hamiltonian in the BSE.
Only a small number of interaction matrix elements (both exchange and direct terms) need to be computed explicitly; all others can be generated by applying symmetry transformations to these elements.
Moreover, transforming the conventional electron-hole product basis to a symmetry-adapted product basis using a projection operator from group theory brings the Hamiltonian 
into a sparse, block-diagonal form, one block per irreducible
representation (irrep).
For the systems considered in this study, only one of the blocks turns
out to contribute to the optical absorption spectrum. The other blocks
do not contribute because the oscillator strengths of their
corresponding eigenstates vanish owing to symmetry.
As a result, the dimension of the electron–hole Hamiltonian for bulk silicon is reduced by a factor of five, leading to a theoretical 125-fold speedup of the leading cubic scaling term during diagonalization. In the case of bulk MoS$_2$, this speedup factor reaches 216.

We also showed how the divergence of the screened Coulomb interaction in reciprocal space 
in the long-wavelength limit $\vbq\rightarrow\mathbf{0}$ can be treated accurately, including the case of anisotropic screening described by a nondiagonal dielectric tensor. Among the systems studied in this work, bulk MoS$_2$ is an example of a material with such a nondiagonal dielectric tensor.
Once a set of eigensolutions is available for the optical
absorption spectrum, determining the corresponding set of eigensolutions for the
electron-energy loss spectrum requires only negligible additional computational effort.

Because of the large size of the electron–hole Hamiltonian, simulations of big systems or calculations with dense $\vbk$-point meshes require the matrix to be distributed across multiple compute nodes.
We therefore developed efficient parallel algorithms (1) to distribute
the Hamiltonian matrix elements -- whose computation is noncontiguous
because of the exploitation of crystal symmetries -- across compute nodes and (2)
to transform the electron-hole Hamiltonian into a symmetry-adapted basis and, at the same time, distribute the matrix into a  block-cyclic layout across the processes,
suitable for the application of diagonalization routines from
ScaLAPACK or ELPA.

The speedup achieved by the use of symmetry enabled BSE calculations for Si to be converged to a very fine $\vbk$-point set of 60$\times$60$\times$60 using a product basis
band cutoff of $n_\mrm{occ}n_\mrm{unocc}=5.6$ (see Sec.~\ref{sec:results}).
The calculated optical absorption spectrum closely reproduces the experimental lineshape and yields an exciton binding energy of 22.5~meV, in better agreement with the experimental value of 14~meV than in previous BSE studies.
Similarly good agreement is obtained for LiF. The optical absorption spectrum of MoS$_2$ was calculated including spin–orbit coupling, yielding the characteristic double-peak structure at the onset of the spectrum. The calculated exciton binding energy of 76~meV is significantly closer to the experimental value of $\sim$85~meV than in previous BSE studies.
The most demanding calculation involved an electron–hole Hamiltonian with a dimension of nearly two million for LiF.



	
	
	\begin{acknowledgments}
We thank Dmitrii Nabok, Francesco Sottile, and Matteo Gatti for fruitful discussions.
Computing time on the supercomputer JURECA~\cite{JURECA} at Forschungszentrum Jülich provided under grant no.~Topomag and cjiff13 is gratefully acknowledged.
This work was supported by the European Centre of Excellence MaX “Materials design at the Exascale” (Grant No. 101093374).
	\end{acknowledgments}

\appendix

\section{Coulomb divergence}\label{sec:appendix}

The Coulomb interaction in reciprocal space exhibits a $q^{-2}$ divergence in the long-wavelength limit, $q\rightarrow 0$. This divergence must not be neglected but instead be carefully integrated over in the evaluation of the interaction matrix elements. In this Appendix, we discuss the treatment of the Coulomb divergence in detail.

According to Eqs.~(\ref{eq:BSE-eigval}) and (\ref{eq:w_interaction}), matrix elements of $w=\overline{v}$ are required only for $\vbq=\vb{0}$, whereas matrix elements of $w=W$ are needed for all elements of the $\vbq$-point set because of the condition $\vbq=\vbk-\vbk'$, and $\vbk'$ runs over all wave vectors. Since $\overline{v}$ is a modified Coulomb potential that specifically lacks the long-range Fourier component, there is no singularity in $\overline{v}$. However, the divergent elements of $W$ do play a role. 
    
In the limit of small momentum transfers $\vbq$, 
the projections $\langle e^{i\vbq\vbr}\varphi_{\vbk n}|\varphi_{\vbk+\vbq n'}\rangle$ of the hole-hole ($n=o$, $n'=o'$) and electron-electron products ($n=u$, $n'=u'$) onto the plane wave $e^{i \vbq \vbr}$ are approximately $1$ if the band indices are equal ($o=o'$ or $u=u'$) (orthonormality) and proportional to $q$ if the band indices differ ($o\neq o'$ or $u\neq u'$) ($\vbk\cdot\vb{p}$ perturbation theory). To be more precise, the proportionality in the latter case has the form $\vb{a}\vbq$ with a vector $\vb{a}$. 
As a consequence, divergent terms proportional to $1/q^2$ formally appear in the diagonal elements of the BSE electron-hole Hamiltonian [Eq.~(\ref{eq:BSE-eigval})] where $u=u^\prime$,  $o=o^\prime$, and  $\vbk=\vbk^\prime$. All other divergent contributions vanish: The wing elements behave as $\vb{b}\hat{\vbq}/q$ (with some other vector $\vb{b}$), which, combined with the diagonal case $o=o'$ and $u=u'$, integrates to zero. Likewise, the combination of $1/q^2$ and $o=o',~u\neq u'$ or $o\neq o',~u=u'$ gives rise to the same kind of divergence $\vb{a}\hat{\vbq}/q$ and thus also integrates to zero.

	
We are thus left with the $1/q^2$ terms in the diagonal matrix elements. Of course, actual singular elements should not appear in the BSE Hamiltonian.
To get an understanding of how to treat the singularity appropriately,
we push the size of the simulation cell, in which periodic boundary
conditions must hold for any finite size, to infinity. As a consequence, the wave vector $\vbk$, which can assume only discrete (vector) values in finite simulation cells, becomes a continuous variable. The infinite simulation cell, together with a continuous wave vector $\vbk$, eventually allows the exciton to be truly localized, and it also resolves the problem of the long-range divergence, because integrating over it in reciprocal space leads to a finite value.
Viewed from a different perspective, when considering a finite $\vbk$-point set, we can imagine the integrand to be interpolated between the discrete $\vbk$ points. This interpolation effectively averages over the divergent part of $W$  and identifies each basis function $\psi_{\vbk uo}$ with a localized wave packet consisting of continuous momenta centered around       each discrete  $\vbk$ point.
The $\vbk'$ summation of Eq.~(\ref{eq:BSE-eigval}), when regarded as an
integration over a wave vector, not only includes the point $\vbk'=\vbk$ ($\vbq=\vb{0}$) but also all points in its neighborhood. In a manner of speaking, the form of the divergence is mapped out in the $\vbk'$ integration. 
    
The divergence is limited to the $\vb{G}=\vb{G}'=\vb{0}$ component of $W$ and has the form
\begin{equation}
    W_{\vb{00}}(\vbq)=\frac{4\pi}{q^2}\,\frac{1}{\hat{\vbq}^\mrm{T}\vb{L}\hat{\vbq}}
    \label{eq:Wdiv}
\end{equation}
around $\vbq=\vb{0}$ with $\hat{\vbq}=\vbq/q$ and a $3\times 3$ tensor
$\vb{L}$ \cite{Friedrich2010,Friedrich21e}. (We note again that $W$ is
$\omega$ dependent in general, and so is $\vb{L}$, but this $\omega$
dependence is omitted because of the static approximation employed in the BSE.)
Clearly, integrating over a $q^{-2}$ divergence in 3D yields a finite value, but the question arises over which region to integrate. To avoid double counting, the integration region should be somehow limited to $\vbq=\vb{0}$ and its neighborhood without including any of the nonzero $\vbq$ points of the $\vbq$ mesh, since each $\vbq$ point possesses its own matrix element in Eq.\ (\ref{eq:BSE-eigval}). 
One might consider integrating over a region stretching halfway to the
neighboring $\vbq$ points. To simplify the integration, one might
furthermore define the integration region to be spherical. However, we
find that such a treatment of the divergence gives rise to poor $\vbq$-point convergence. Furthermore, it would not account for anisotropic (e.g., layered) systems or unevenly distributed $\vbq$-point meshes.

We employ the same treatment of the divergence as the one introduced in Ref.~\onlinecite{Friedrich2010} for the $GW$ method. Here, we describe an extended version that allows for anisotropy in the screening, i.e., tensors $\vb{L}$ that are not just a multiple of the identity matrix. 
The basic idea is that the region of integration is extended over the
whole reciprocal space. As this integration comprises all other $\vbq$
points, too, we have to introduce a double-counting correction.
Another difficulty is that $q^{-2}$ does not integrate to a finite
value if  the integration extends over the whole space. We address this difficulty by replacing $1/q^2$ in Eq.\ (\ref{eq:Wdiv}) by $e^{-bq^2}/q^2$ with a small parameter $b>0$. This is a modification of an idea introduced in Ref.~\onlinecite{Massidda93} in the context of the Hartree-Fock method. The exponential factor guarantees that both the integral and $\vbq$ sum remain finite in the contribution of the singularity,
\begin{equation}
    W_{\vbk n,\vbk n';
    \vbk n,\vbk n'}^\mrm{div}=
    \frac{1}{2\pi^2}\int d^3 q \frac{e^{-bq^2}}{q^2\hat{\vbq}^\mrm{T}\vb{L}\hat{\vbq}}-
    \frac{4\pi}{N_\vbk V_\mrm{uc}}\sum_{q\neq 0}
    \frac{e^{-bq^2}}{q^2\hat{\vbq}^\mrm{T}\vb{L}\hat{\vbq}} ,
\end{equation}
to the diagonal terms of Eq.\ (\ref{eq:w_interaction2}) for $w=W$, where the sum over $\vbq$ is the double-counting correction. To evaluate the integral, we first consider the angular integration. The only angle-dependent term is $\hat{\vbq}^\mrm{T}\vb{L}\hat{\vbq}$. Its expansion in terms of spherical harmonics can be written as 
\begin{equation}
\hat{\vbq}^\mrm{T}\vb{L}\hat{\vbq}=\sum_{l=0,2}\sum_{m=-l}^{l}L_{lm}Y_{lm}(\hat{\vbq})\ .
\label{eq:L0}
\end{equation}
Likewise, we make the general ansatz
\begin{equation}
\frac{1}{\hat{\vbq}^\mrm{T}\vb{L}\hat{\vbq}}=\sum_{l=0,2,4,\dots}\sum_{m=-l}^{l}H_{lm}Y_{lm}(\hat{\vbq})
\label{eq:hinv0}
\end{equation}
for its reciprocal with unknown coefficients $H_{lm}$. Multiplication and using the Gaunt coefficients $G_{LM,lm,l'm'}=\int 
Y^*_{LM}(\hat{\vbq})
Y_{lm}(\hat{\vbq})
Y_{l'm'}(\hat{\vbq})
d^2\hat{\vbq}$ gives
\begin{align}
\nonumber
1=&\sum_{L=0,2,4,\dots}\sum_{M=-L}^{L}
\left[
\sum_{l=0,2}\sum_{m=-l}^{l}
\sum_{l'=|L-l|}^{L+l}
G_{LM,lm,l',M-m} \right. \\
&\times \left. L_{lm}H_{l',M-m}
\vphantom{\sum_{l'=|L-l|}^{L+l}}
\right]
Y_{LM}(\hat{\vbq})\ ,
\label{eq:hinv1}
\end{align}
which, by equating the coefficients on both sides [note that $\sqrt{4\pi}Y_{00}(\hat{\vbq})=1$], yields a system of linear equations.  The upper bound of the $l$ sum ($L$ sum) in Eq.\ (\ref{eq:hinv0}) [Eq.\ (\ref{eq:hinv1})] is a convergence parameter. Since all $Y_{lm}(\hat{\vbq})$ integrate to zero on the unit sphere except $Y_{00}(\hat{\vbq})$, it is sufficient to converge only the coefficient $H_{00}$ with respect to the summation upper bound. We find that convergence is very fast. Using the spherical-harmonic expansions of Eqs.~(\ref{eq:L0}) and (\ref{eq:hinv0}) generally accounts for anisotropic dielectric screening.

We finally get
\begin{equation}
    \int_{q<q_\mrm{max}} d^3q \frac{e^{-bq^2}}{q^2\hat{\vbq}^\mrm{T}\vb{L}\hat{\vbq}}
    =
    \pi\frac{H_{00}}{\sqrt{b}}\mrm{erf}(\sqrt{b}\,q_\mrm{max})
    \,.
    \label{eq:divergence_int}
\end{equation}
with the error function $\mrm{erf}(x)=(2/\sqrt{\pi})\int_0^x e^{-x'^2}dx'$.
We have introduced a cutoff radius $q_\mrm{max}$ up to which the $\vbq$ integration and summation are carried out. The reason for introducing $q_\mrm{max}$ is that, even with the exponential factor, the addends do not fall off quickly enough to enable a fast evaluation of the $\vbq$ sum. In addition, we smoothen the upper limit of the sum by an additional function
\begin{equation}
\sum_{\substack{q\neq 0\\q<q_\mrm{max}-\Delta}}
    \frac{e^{-bq^2}}{q^2\hat{\vbq}^\mrm{T}\vb{L}\hat{\vbq}}\,
+
\sum_{\substack{q\neq 0\\|q-q_\mrm{max}|\le\Delta}}
    \frac{e^{-bq^2}}{q^2\hat{\vbq}^\mrm{T}\vb{L}\hat{\vbq}}\,f\left(\frac{q-q_\mrm{max}}{\Delta}\right)
    \label{eq:divergence_sum}
\end{equation}
with $\Delta=(V_\mrm{BZ}/N_\vbk)^{1/3}$, the Brillouin-zone volume $V_\mrm{BZ}=8\pi^3/V_\mrm{uc}$, and $f(x)=(x^3-3x+2)/4$.
To minimize the number of parameters, $b$ and $q_\mrm{max}$ are linked by the condition $b=e^{-bq_\mrm{max}^2}$. We have found $b=0.005$ to be a good choice.
 
	\bibliography{main}
	
\end{document}